\documentclass{elsart}

\usepackage{graphicx}
\usepackage{textcomp}
\usepackage{amssymb}
\newcommand{\mod}[1]{\textrm{#1}}   

\begin{document}

\begin{frontmatter}



\title{Confocal Microscopy of Colloidal Particles: Towards Reliable, Optimum Coordinates}


\author[Edinburgh,DDorf]{M.C. Jenkins\corauthref{cor}},
\corauth[cor]{Corresponding author. Tel.+49-211-81-14337, fax.
+49-211-81-14850} \ead{matthew.jenkins@uni-duesseldorf.de}
\author[Edinburgh,DDorf]{S.U. Egelhaaf}
\ead{stefan.egelhaaf@uni-duesseldorf.de}

\address[Edinburgh]{Scottish Universities Physics Alliance (SUPA),
Collaborative Optical Spectroscopy Micromanipulation and Imaging
Centre (COSMIC), and School of Physics, The University of Edinburgh,
Kings Buildings, Mayfield Road, Edinburgh, EH9 3JZ, United Kingdom}

\address[DDorf]{Condensed Matter Physics Laboratory, Lehrstuhl f\"{u}r
Physik der weichen Materie, Heinrich-Heine-Universit\"{a}t
D\"{u}sseldorf, Universit\"{a}tstra{\ss}e 1, D-40225 D\"{u}sseldorf,
Germany}

\begin{abstract}
Over the last decade, the light microscope has become increasingly
useful as a quantitative tool for studying colloidal systems.  The
ability to obtain particle coordinates in bulk samples from
micrographs is particularly appealing.  In this paper we review and
extend methods for optimal image formation of colloidal samples,
which is vital for particle coordinates of the highest accuracy, and
for extracting the most reliable coordinates from these images. We
discuss in depth the accuracy of the coordinates, which is sensitive
to the details of the colloidal system and the imaging system.
Moreover, this accuracy can vary between particles, particularly in
dense systems. We introduce a previously unreported error estimate
and use it to develop an iterative method for finding particle
coordinates.  This individual-particle accuracy assessment also
allows comparison between particle locations obtained from different
experiments. Though aimed primarily at confocal microscopy studies
of colloidal systems, the methods outlined here should transfer
readily to many other feature extraction problems, especially where
features may overlap one another.
\end{abstract}

\end{frontmatter}

\tableofcontents

\section{Introduction}
Confocal microscopy has made an enormous impression on the
biological sciences \cite{PawleyConfocal}, and started to do the
same in colloidal science.  In particular, the ability to perform a
quantitative analysis, especially to extract particle coordinates
from genuinely bulk colloidal samples, is hugely useful in providing
information at the single particle level, and is ideal for
comparison with the results of computer simulations. Whilst the
range of experimental avenues to be explored is vast, several themes
have emerged. Paradoxically, arguably more difficult systems such as
glasses and gels have been studied in detail, as we discuss below.
This is partly because of present outstanding questions in these
which are likely only to be answered with knowledge of localised
events, which more traditional `averaging' techniques such as light
scattering cannot easily access, but also because more tightly
confined systems move at rates better suited, i.e. more slowly, to
the more sedate hardware available until very recently.

Amongst the very early studies of colloidal samples was a structural
analysis of a very high density glass \cite{vanBlaaderen95}.  This
lead to studies of (particularly heterogeneous) dynamics in
supercooled liquids and glasses
\cite{Weeks00,Weeks02b,Kegel00b,Gasser03,Conrad05}.  Similar studies
concentrated on aging
\cite{Courtland03,Cianci06,Cianci06b,Simeonova04} and the response
of samples to localised perturbation \cite{Habdas04}. Recently, in
response to the experimental verification by light scattering of a
re-entrant glass transition \cite{Pham02}, the effect of an
increasing short-range particle attraction on the glass transition
has been studied \cite{Simeonova06}.  In a similar fashion but for
different attractions, other groups have studied gels
\cite{PSmithThesis,Dinsmore02,Varadan03,Gao04,Sanchez05,Dinsmore06,Dibble06}.

As well as disordered systems, there have been studies of bulk
crystals in systems with both hard sphere \cite{Campbell02} and more
complex interactions
\cite{Royall03,Yethiraj03,Leunissen05,Campbell05}, as well as
crystallisation \cite{Gasser01,Hoog01}, stacking disorders
\cite{Verhaegh94}, and growth (including epitaxial) of crystals
\cite{vanBlaaderen97b,vanBlaaderen97c,vanBlaaderen03,Hoogenboom03}.
Though many of these take advantage of the particle coordinates for
structural and dynamic analyses, it is also possible to infer
thermodynamic properties directly by microscopy \cite{Dullens06b}.

Recently, interest has developed in the behaviour of glasses under
shear \cite{Derks04,Besseling06} and flow \cite{Isa06}. Studies
such as these benefit enormously from the advent of video-rate and
faster confocal microscopes.  At the other temporal extreme, studies
on force distributions in emulsions
\cite{BrujicThesis,Brujic03,Brujic03b} and sedimenting
nearly-colloidal systems \cite{mythesis,mybridgespaper1}, in which
particles are forced into intimate contact, present equal
challenges.

Several papers have described microscopy of colloids, both for video
and confocal microscopy (reviewed in \cite{Murray96} and
\cite{Chestnut97} respectively).  The introductions by Weeks
(\cite{WeeksWeb}, and latterly \cite{SemwogerereBook}) are
accessible general references.  The standard reference for
quantitative studies, i.e. particle location and tracking,
\cite{Crocker96}, is essentially for two dimensional studies using
video microscopy. Others extended these to three dimensional
confocal microscopy \cite{Dinsmore01,Habdas02,Prasad07}.  Some
details of the other groups' techniques have emerged
\cite{Royall03,Bromley01} and an excellent reference is web-based
\cite{WeeksTrackingWeb}.

The above examples encompass a huge range of typical particle types,
configurations, and imaging conditions. Key parameters which vary
widely include the volume fraction (ranging at least from
$\Phi{=}0.002$ to $\Phi{\simeq}0.64$ \cite{Yethiraj03,mythesis});
particle size; typical particle separation (very different for gels
and attractive glasses from that for coexistence crystals, even at
the same volume fraction); sample composition (especially refractive
index mismatch); the nature of the particles themselves
(particularly whether the whole particle is visible to the imaging
system); and the speed at which images can be captured. All of these
affect the success of particle location, but these issues are not
thoroughly considered in many papers. We emphasise particularly that
studies which claim structural differences between different systems
(e.g. \cite{Conrad05}) should justify that the particle location
works equally well in both systems.

Our aim is for {\em reliable} and {\em optimum} coordinates. By
optimum we mean both that the coordinates extracted from a given
data set are the best that can be inferred from that data set, and
that the data set contains the greatest possible information about
the sample. Crucial to {\em optimum} coordinates is therefore
optimum image capture, and optimum coordinate extraction, both of
which we discuss here.  Reliable refers to the fact that we have
confidence --- that is, an objective quantification --- in each
coordinate measured.  Also in this paper, we introduce a means of
comparing accuracy between data sets of differing quality; this is a
measure of reliability. With this, we are able to obtain particle
coordinates in difficult (what is meant by `difficult' will become
clear) samples, where, although the accuracy is necessarily low, we
can nonetheless reliably locate the particles. Crucially, we argue
that if meaningful quantitative comparisons are to be made between
different systems, it is vital that such a measure is available.
Quantitative confocal microscopy of colloids works well under
favourable conditions (see the above citations), and we describe
here how one can know how successful it has been.

Although we are interested in colloidal systems, we note that much
of what we discuss has been considered in other fields.  Some
prominent examples are particle image velocimetry (PIV), biophysics
and astronomy.  PIV does not quite track particles, but it seeks the
location of correlation peaks using techniques similar to ours
\cite{Raffel98}. In biophysics, tracking of fluorescently dyed
objects, for example proteins and lipids, as they diffuse on cell
surfaces is a common problem \cite{Thompson02,Cherry98,Saxton97} and
astronomers have developed many advanced algorithms, in particular
for detecting x-ray sources, for example \cite{Ayres04,Feigelson92}.
There is inevitable overlap between these fields, since they all
seek to locate features with maximum precision.  Doubtless there is
scope for further consolidation between these fields, but equally
there is a strong motivation for discussing the particular
requirements of colloidal particle location in the microscope.
Nonetheless, the contents of this paper should apply outside of
colloidal science, and in particular need not be limited to confocal
studies; many bright features can be located using the techniques we
describe.

This paper covers the whole process of particle location, from the
imaging of objects in the colloidal regime, to testing the validity
of the extracted coordinates. In section 2 we discuss imaging of
colloid-sized objects.  We elaborate in section 3 on practical
confocal microscopy of colloidal systems.  Section
\ref{findingcoords} discusses strategies for coordinate extraction,
and Section \ref{accuracychecks} describes some basic tests of
accuracy. Section \ref{centroiding} discusses the centroiding
technique in some detail.  Thereafter, in Section 7, we describe our
improvements on this method, including a detailed error analysis,
its success, and limitations.


\section{Imaging Colloid-sized objects}
In this Section, we describe the confocal micrograph of a spherical
colloid-sized object, which is subtly distorted by the imaging
system.  The distortion is an inherent feature of imaging objects
near to the diffraction limit of any imaging system.  Its nature is
important when attempting to locate objects in the colloidal size
range, particularly in dense systems. For spheres, the micrograph
which the imaging system produces is the sphere spread function
(SSF).  It will be described in detail, as will noise, which must be
carefully dealt with to obtain optimum coordinates.

\subsection{Imaging Process}\label{ImagingProcess}
\begin{figure}[h]
\begin{center}
\includegraphics[width=10cm,bb=0 0 600 286]{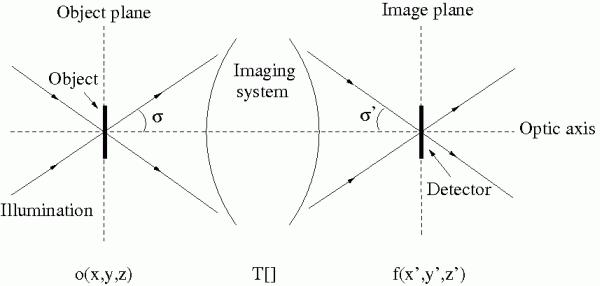}
\caption[The generic imaging process.]{The generic imaging process,
comprising a radiating object $o(x,y,z)$, an imaging system,
$T$[$\:$], and the detected image,
$f(x',y',z')$.\label{genericimaging}}
\end{center}
\end{figure}

Figure \ref{genericimaging} shows the generic imaging process:
visible electromagnetic radiation emanating from the object is
represented by $o(x,y,z)$, whilst $f(x',y',z')$ is the image. The
operation which transforms one into the other is performed by the
imaging system and is designated $T[\:]$:
\[f(x^\prime,y^\prime,z^\prime)=T[o(x,y,z)].\]


An extended object $o(x,y,z)$ can be considered a series of points
on a finely-spaced lattice. An arbitrary point $(i,j,k)$ in this
lattice is specified by the delta function
$\delta(i{-}x,j{-}y,k{-}z)$, allowing the object to be rewritten in
terms of these lattice coordinates:
\[o(i,j,k)={\int}{\int}{\int} o(x,y,z)\delta(i{-}x,j{-}y,k{-}z)dxdydz,
\]
and where the integrals are over the field of view. This is a null
operation, but permits us to proceed.  The imaging process $T[\:]$
can be written as a convolution, a description which is useful since
both digital filtering and modelling the image of a spherical
colloidal particle are performed in these terms.  Now, an image is
given by $f(x^\prime,y^\prime,z^\prime)=T[o(x,y,z)]$, since the
coordinate system used to describe the object cannot be important,
so that:

\[f(x',y',z')=T\left[ {\int}{\int}{\int} o(x,y,z)\delta(x'{-}x,y'{-}y,z'{-}z)dxdydz \right],\]
which, provided the imaging operation $T[\;]$ is linear, reduces to:
\begin{equation}\label{convolutionintegral}
f(x',y',z')={\int}{\int}{\int}
o(x,y,z)T[\delta(x'{-}x,y'{-}y,z'{-}z)]dxdydz.\end{equation} From
this we identify the distribution
$T[\delta(x'{-}x,y'{-}y,z'{-}z)]{=}p(x,y,z)$ as the image of a
single point, which in any real imaging system is not itself a
single point; it is the {\bf point spread function} ({\bf PSF})
whose origin is the diffraction limit of the imaging system, and
which is widely described in standard texts, for example
\cite{Hecht}.

The expression for the observed intensity, $f(x',y',z')$, is a
convolution integral; $f(x',y',z')$ is the convolution of the system
PSF with the object illuminance, which can be written as
\begin{equation}\label{eqPSF}
f(x',y',z')=T[\delta(x'{-}x,y'{-}y,z'{-}z)]\odot o(x,y,z)=p(x,y,z)\odot o(x,y,z),
\end{equation}
where `$\odot$' denotes convolution.

The image produced by an imaging system with linear $T$ is simply
the convolution of the effect that it has on a single point object
with the original image; the imaging system places a copy of the
point spread function at each point in the image, scaled by the
intensity of the object at the corresponding point.

It is well documented that the PSF for the light microscope is well
modelled by a Gaussian in both the lateral and axial directions
(e.g. \cite{mythesis,Hecht,MSE,WilsonConfocal}). Also for other
imaging systems it is possible to model accurately the PSF, and
there are commercial software packages which can do this (for
example, Huygens and Autoquant,
http://www.svi.nl/products/professional/ and
http://www.aqi.com/index.asp respectively). Alternatively, one could
measure the response of sub-microscopic, essentially `point'
objects, many times to obtain a convincing (mean) system PSF.
However, for our purposes, since the first peak in the PSF contains
overwhelmingly the majority of the intensity of the distribution,
and since the central peak is well modelled by Gaussians, the system
PSF can be modelled by a simple set of Gaussians placed in a single
array.
\begin{displaymath}
i(x,y,z)=i_0exp[-\frac{1}{k_1}\left[ (x{-}x_0)^2 + (y{-}y_0)^2
\right] - \frac{1}{k_2}(z{-}z_0)^2]
\end{displaymath}
populated the array in one calculation, and allowed the relative
lateral and axial extents to be varied. The extent (the variance,
$\sigma^2{\equiv}\frac{k}{2}$) of the PSF was adjusted via $k_1$ and
$k_2$, to correspond to reasonable values for the confocal
microscope.

\subsection{Imaging a Colloid-sized Object \label{comparisonSSFmodelledreal}}
Every imaging system distorts the object by the PSF.  In the worst
cases, this renders the object unresolvable.  In contrast, if the objects
under study are much larger than the resolution limit, the
distortions are not noticed (photographs, television images).  In
imaging colloids, we investigate an intermediate situation where
diffraction effects are important but not overwhelming.

\begin{figure}[h]
\begin{center}
\includegraphics[width=10cm,bb=0 0 800 245]{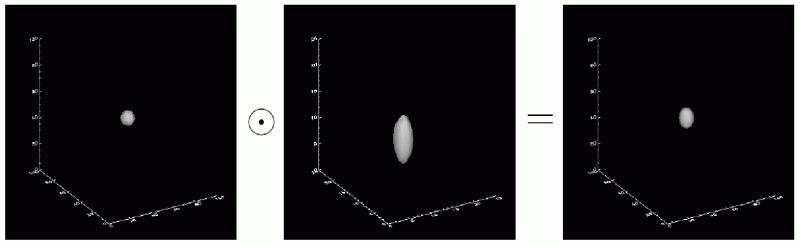}
\end{center}
\caption{Model of a homogeneously-dyed spherical particle of radius
$0.5$ $\mu m$ $i_{sphere}(x,y,z)$ (left), the PSF $p(x,y,z)$
(centre) and the corresponding image $s_{model}(x,y,z)$
(right).\label{Model1} }
\end{figure}

Following the discussion above, we recognise the sphere spread
function (SSF) as the convolution of true sphere response,
$i_{sphere}(x,y,z)$, with the appropriate PSF:
\begin{equation}\label{SSFeq}
s_{\mathrm{model}}(x,y,z)=i_{sphere}(x,y,z)\odot p(x,y,z)
\end{equation}
We describe a simple model of this process for an indication of the
resultant SSF. We will use this to justify our particle location
technique. In particular, we consider an approximation to the
fluorescence confocal image of a homogeneously-dyed spherical
colloidal particle. (The scattering of light from a dielectric
sphere is more complicated \cite{Kerker69} and does usually not give
rise to a uniform intensity over the sphere, but this is beyond the
scope of the present article.) We convolve its `true' image
$i_{sphere}(x,y,z)$ with the PSF of the imaging system, $p(x,y,z)$,
(Equation \ref{SSFeq} and Figure \ref{Model1}). Intensity
distributions through the resultant modelled SSF compare well with
confocal images of homogeneously-dyed PMMA particles.  For more
details, consult \cite{mythesis}.


The PSF of a point particle is well modelled by a Gaussian surface
of revolution. This is not necessarily the case for larger particles
and thus suggests that a Gaussian does not always model the SSF; its
success is likely to vary with the size of the spheres.  Although a
Gaussian approximation does account for the important properties
quite well (the shape is roughly right, and it is spherically
symmetric) \cite{Crocker96}, it has been pointed out recently that
the lobes can have an important effect on the inferred particle
locations \cite{Baumgartl05}.

Even with our simple Gaussian approximation, we see the important
feature: the `smearing' of a micron sized sphere due to the
diffraction limit results in an SSF that is larger in extent than
the true size of the sphere. In the case where the spheres are well
separated, this is of no consequence.  However, when the spheres are
close to one another, the SSFs overlap, a detail which turns out to
be crucial in microscopy of some colloidal samples, and which we
discuss in detail later.

We argued above that the image of a particle can be written as a
convolution of its illuminance $i(x,y,z)$ with the imaging system
PSF $p(x,y,z)$, equation~\ref{SSFeq}.  In a
similar way, we can view the placement of this image at any
particular point in space (i.e. the particle coordinate) as a
convolution with a Dirac delta function at that point.
\begin{equation}\label{eqSSF}
f(x',y',z')=\delta(\mathbf{r})\odot s(x,y,z),
\end{equation}
where $s(x,y,z)$ represents the sphere spread function (modelled,
such as $s_{\mathrm{model}}(x,y,z)$ above, or measured) which
corresponds to a sphere centred on the origin (thereby defining
implicitly an origin).

By the convolution theorem, we can write this as
\begin{equation}
\hat{F}(u,v,w)=\mathcal{F} \{ \delta(\mathbf{r}) \} \times
\hat{S}=\alpha \hat{S},\label{eqnSSFconv}
\end{equation}
where $\hat{F}(u,v,w)$ is the Fourier Transform of the image
$f(x',y',z')$, $\hat{S}$ is that of the SSF $s(x,y,z)$ and
$\mathcal{F} \{ \delta(\mathbf{r}) \}$ is that of a delta function
at the particle's position $\mathbf{r}$.  The last equality follows
since the Fourier transform of a delta function centred on the
origin is simply a constant ($=1/\sqrt{2\pi}$ or $1$, depending on
the definition), $\mathcal{F} \{ \delta(\mathbf{r}) \}\equiv
\alpha$. This idea of the image of a particle as an instance of a
motif will be useful later.

\begin{figure}[h]
\begin{center}
\includegraphics[width=10cm,bb=0 0 800 486]{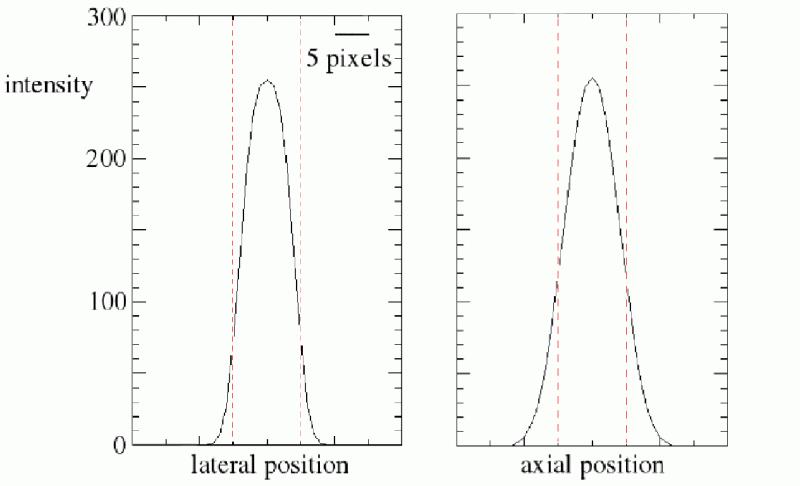}
\caption[The simulated intensity profile through a single
particle.]{Simulated intensity profile of a particle with $2$ $\mu$m
diameter in the lateral (left) and axial (right) direction. The true
size of the particles ($10$ pixels) is indicated by the dashed
lines. The axes are in pixels, with a pixel pitch of $0.2$ $
\mu$m.\label{singleparticle}}
\end{center}
\end{figure}

The intensity profile through a typical modelled SSF in lateral
(left) and axial (right) direction are shown in
figure~\ref{singleparticle}. This SSF represents a sphere with a
diameter of $2$ $\mu m$ and was calculated using a Gaussian PSF of
lateral extent $300$~nm and axial extent $600$~nm. The centroid of
the brightness within a window about the particle corresponds to the
centre of the particle (within uncertainty).

We have seen that the image of an object is its true form convolved
with the instrument resolution function (the PSF), and how this
affects the optical micrograph of a colloidal particle. The degree
of broadening is purely a function of the system PSF, so one readily
appreciates that the PSF would ideally be itself a point. Though
diffraction means that this is never true, it can be made more
nearly so. One widely used means of narrowing the PSF is the laser
scanning confocal microscope (LSCM), which we consider now.

\subsection{Confocal Microscopy of Colloid-sized Objects}
Instrumental (PSF) related `smearing' always hampers
imaging colloid-sized objects with the light microscope, but is
particularly compromising for bulk samples. There have been
successful bulk colloidal studies using a conventional optical
microscope \cite{MSE,Elliot97,Elliot01,Elliot01b}, but the large
extent (particularly axially) of the PSF is problematic.  The
confocal microscope addresses this difficulty particularly well.

\begin{figure}[h]
\begin{center}
\includegraphics[width=10cm,bb=0 0 600 197]{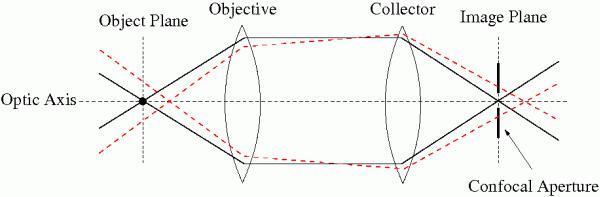}
\caption[The confocal principle.]{The confocal principle: light
originating from within the focal plane (solid black) is imaged at
the image plane, as in a conventional microscope. Light originating
outwith the focal plane (dashed red) is not brought to a focus at
the image plane, and appears as blur in a conventional microscope. A
confocal aperture in the image plane ensures only a thin optical
section is imaged, so that this detrimental effect is eliminated by
the confocal microscope.\label{confocalprinciplediag}}
\end{center}
\end{figure}

Figure \ref{confocalprinciplediag} shows a simple representation of
the confocal principle (for technical details see, e.g.
\cite{PawleyConfocal,mythesis,WeeksWeb,SemwogerereBook,WilsonConfocal}).
The two sets of rays illustrate imaging of a point at the focal
plane (solid black) and one outwith the focal plane (dashed red).
Both are brought to focus at different points; light originating
from outwith the object focal plane appears in the detected image as
out-of-focus blur. The image is substantially improved by
eliminating this by inserting an aperture at the detector
(illustrated); the image plane is {\bf con}jugate with the {\bf
focal} plane, hence {\bf confocal}. The pinhole size is a compromise
between the greater confocal effect of a small pinhole and the
reduced light budget this affords. Due to the difficulty of
manufacturing small variable apertures the pinhole size is not
usually continuously variable. A near-match is usually more than
satisfactory. The confocal aperture reduces the axial extent of the
PSF, and whilst this is its most dramatic influence, it also
diminishes the PSF lateral extent \cite{mythesis,WilsonConfocal}.
The confocal microscope is thus able to provide finely resolved
images from within a bulk sample: this optical sectioning is
overwhelmingly the most important benefit of the confocal microscope
for colloidal studies. This improvement in resolution is at the
expense of its field of view, which can subsequently be recovered by
scanning \cite{Lukosz66}.

\subsection{Noise \label{noise}}
So far we have assumed that the detector accurately reproduces the
response of the object under study.  In a real experiment, however,
there is also {\bf noise}, an all-inclusive term for processes which
affect the image but which do not relate to the object itself.
Examples include the discrete nature of the radiation; varying
detector sensitivity; unstable illumination; electrical noise; and
transmission errors.

The {\bf signal-to-noise ratio (SNR)} of an image is a means of
quantifying the extent of noise-derived degradation
\cite{GonzalezWoods}. No one measure of SNR fully quantifies the
`quality' of an image; most image processing addresses images which
are ultimately for (inherently subjective) human observation. We
give one definition of SNR:

If the detected image $g(i,j,k)$ comprises a `signal' part $f(i,j,k)$ and a
`noise' part $n(i,j,k)$:
\begin{equation}\label{eqNoise}
g(i,j,k)=f(i,j,k)+n(i,j,k)
\end{equation}
then the respective variances are $\sigma^2_f=\langle
|f(i,j,k){-}\langle f(i,j,k)\rangle|^2\rangle$ and
$\sigma^2_n=\langle |n(i,j,k)|^2\rangle$. The SNR is
simply defined as:
\[\mathrm{SNR}=\frac{\sigma_f}{\sigma_n}=\sqrt{\frac{\sigma_g^2}{\sigma_n^2}-1},\]
where the second equality follows since the noise in this case is
uncorrelated: $\sigma_g^2=\sigma_f^2+\sigma_n^2$.

To find the SNR, therefore, we require two of $\sigma_n, \sigma_f,
\sigma_g$.  From a single image, we only know $\sigma_g$.
Occasionally, it is possible to extract a region of constant
intensity (a region of sky within a photograph) and from this infer
$\sigma_n$.  Generally, and particularly in colloidal samples, this
is not so, and we must obtain two images ($g_1(i,j,k)$ and $g_2(i,j,k)$)
of the same scene $f(i,j,k)$:
\begin{eqnarray*}
g_1(i,j,k)&=&f(i,j,k)+n_1(i,j,k) \\
g_2(i,j,k)&=&f(i,j,k)+n_2(i,j,k).
\end{eqnarray*}
Taking the normalised correlation between the two realisations
$g_1(i,j,k)$ and $g_2(i,j,k)$:
\[r=\frac{\langle(g_1 g_2{-}\langle g_1\rangle\langle g_2\rangle)\rangle}{\left[\langle|g_1{-}\langle g_1\rangle|^2\rangle\langle|g_2{-}\langle g_2\rangle|^2\right]^{1/2}}
=\frac{\sigma_f^2}{\sigma_f^2+\sigma_n^2}.\]
This permits direct calculation of the SNR
\[\mathrm{SNR}=\sqrt{\frac{r}{1-r}}.\]
With colloidal samples it is, however, seldom possible to form two
successive images of the identical image, so that the SNR can
rarely be calculated for a genuinely representative sample.

Although the SNR is not definitive, we have found that an experienced eye is
satisfactory in determining pre-analysis whether images are
suitable.  This is not crucial, since --- with the aid of the analysis
described in this paper --- we assess the accuracy of individual
coordinates post-processing.

\section{Imaging Bulk Colloidal Samples}
In the previous section, we described how colloidal objects can in
principle be imaged using the confocal microscope. In this section,
we discuss how this is achieved in practice, particularly with
regard to achieving the best possible images.

\subsection{Sample Requirements \label{Nyquist}}
For our analysis, it is required that fluorescence is present in the
system, the particles are of suitable size and
the refractive indices of the particles and suspension medium are
suitably close.

\begin{figure}[h]
\begin{center}
\includegraphics[width=6cm,angle=0,bb=0 0 400 298]{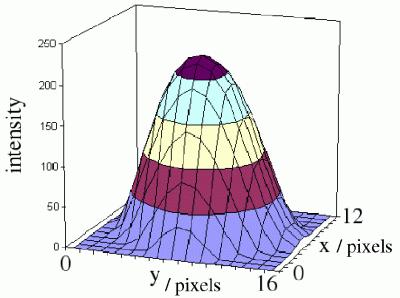}
\caption{Intensity profile through an undyed PMMA sphere dispersed
in an aqueous solution of EOSIN Y.\label{fluorescentsolvent}}
\end{center}
\end{figure}

Most confocal microscopes require fluorescence. For particle
location, it is necessary that particles are {\bf labelled} in order
to distinguish them from the suspension medium. The whole particle
or only parts, e.g. the core, can be dyed. Usually the dye is
chemically attached to the particles, hence preventing dye from
leaching into the solvent. However, fluorescent solvent may be used
with undyed particles (Figure \ref{fluorescentsolvent}). Dye
properties vary in both fluorescence yield and how quickly their
fluorescence property is lost on excitation (photobleaching)
\cite{Campbell02}.


The acceptable size range for the particles is specified by the
resolution of the microscope; around $200$ -- $300$ nm laterally
\mod{and $500$ -- $600$ nm axially} \cite{WilsonConfocal}. According
to the usual Nyquist-Shannon sampling requirement
\cite{Nyquist28,Shannon49}, the distance represented by each pixel
(the pixel pitch) should be around half this. Nevertheless, a
certain degree of undersampling is permissible, since the shape of
the particles is known.  The procedure for locating particles
requires around $10$ pixels per particle (see below), which implies
a particle size of at least around $1$ $\mu$m diameter with larger
particles resulting in more reliable results. Most studies have used
particles with a diameter around $2$ $\mu$m.

The refractive index mismatch between the particles and suspension
medium must be low to minimise the effect of scattering; scattering
events between the entry point of the beam and the position being
imaged reduce image quality. The refractive index mismatch, as well
as the particle concentration and particle size, determine the depth
into the sample where images of sufficiently high quality can be
obtained. Increasing refractive index mismatch, increasing particle
concentration, reducing particle size and increasing depth all
worsen the reliability of particle location. \mod{As an example,
poly-methyl-methacrylate (PMMA) particles labelled with the dye
4-methyl-aminoethylmethacrylate-7-nitro-benzo-2-oxal,3-diazol (NBD)
having} a diameter of about $2~\mu$m and a refractive index
$n{\simeq}1.49$ dispersed in {\em cis}-decalin ($n{\simeq}1.48$)
provide usable images for very dense samples ($\Phi {\simeq} 0.60$)
to a depth of around $50~\mu$m, whereas the same system but with a
very nearly refractive index-matching mixture of {\em cis}-decalin
and cycloheptylbromide (CHB) provides images at more than $100~\mu$m
into the sample (\mod{when the objective lens working distance is
often limiting}). At the other extreme, for silica ($n{\simeq}1.50$)
in water ($n{=}1.33$) useful data is likely to come only from within
a few micrometers of the cover slip.

\subsection{Sample Containers\label{samplecellssection}}
The sample cell significantly contributes to the image quality which
can be achieved from bulk colloidal systems. Although the typical sample
cell is a simple chamber, it must be optically suitable, straightforwardly
filled and then sealed airtight, as well as impervious to the
constituents of the sample.

The optical requirements of the cell for light microscopy have been
discussed in detail before \cite{MSE,Elliot01}.  These are
simplified for confocal studies, since here only one surface of the
cell is in the optical path. The most important requirement is that
the thickness of this cell wall must not exceed the objective
working distance by more than the depth to which one wishes to
observe.

\begin{figure}[h]
\begin{center}
\includegraphics[width=13cm,bb=0 0 800 360]{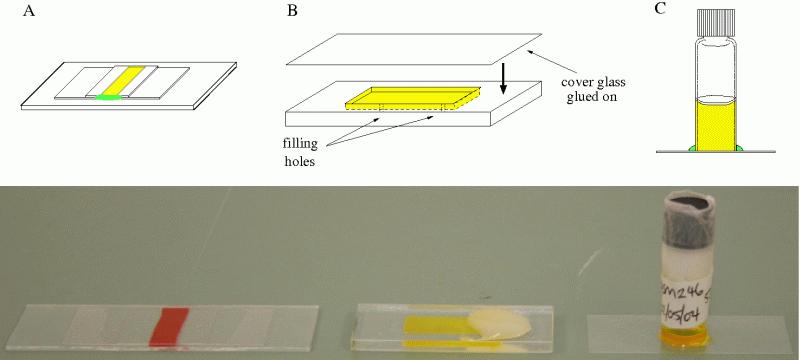}
\caption{Three sample cell constructions. See text for details.
\label{samplecells}}
\end{center}
\end{figure}

The widely-used capillary tubes \cite{MSE,Elliot01} with typical
wall thicknesses of about $100~\mu$m, although very convenient,
provide poor images (which may in part be due to their method of
manufacture). We thus prefer to make use of cover glasses, which are
by design of high optical quality.  Constructing a chamber from
these is generally straightforward; figure \ref{samplecells} shows
three arrangements. The first (Fig.~\ref{samplecells}A) is an
arrangement of cover glasses and a microscope slide which is held
together by use of UV-curing adhesive (e.g. Norland Optical
Adhesive, NOA 61). The number of cover glasses can be varied to form
different sized cavities. The cavities must be filled by capillary
action, so are not suitable for dense and/or viscous samples. A
second cell (Fig.~\ref{samplecells}B) comprises a cover slip
attached to a machined block of material (such as PMMA).  The outer
dimensions of the cell are typically chosen for convenience to match
cover glass sizes, whilst the inner are freely chosen. The cell is
filled through a small hole and air correspondingly evacuated
through a second hole ($0.6$~mm diameter is convenient). Once the
cell is filled, the holes are sealed using UV-curing glue. If
desired, the cell can be made suitable for observation from either
side by attaching a cover glass to one side and a microscope slide
to the other.  The second is by far the most versatile, and remains
sealed for a very long time.  As with the first cell, it is
difficult to fill with the most dense and/or viscous samples. In
particular, care must be taken of the so-called `self-filtration'
effect \cite{MarkHaw04}. The third cell (Fig.~\ref{samplecells}C)
consists of a glass vial with its base replaced by a cover glass
(UV-curing adhesive is again suitable). This cell is easily
 filled even with the most dense samples and its contents can be reused or adjusted, but in our experience cells of
this type are frequently not airtight for long times.

As well optically suitable and impervious to the sample, the
surfaces of the cell should be chosen with consideration \mod{in
view of wall effects.  This depends on the specific project, for
example,} smoothness can cause wall-induced ordering. This can be
avoided by treating the surface. The simplest method is to deposit a
drop of a dilute particle suspension on the (tilted) glass surface
and then bake this once dry (for PMMA, 50 min at 85\textcelsius~is
suitable). Particularly \mod{if polydisperse particles are used,
this provides a good non-slip coating \cite{Besseling06} which
discourages epitaxial crystal growth.} Alternatively, one can
spin-coat the glass surfaces with a layer of PMMA, which is
subsequently covered with a stabiliser such as
poly(12-hydroxystearic acid) (PHS), which is also used to sterically
stabilise particles. The PMMA layer can be omitted by using a PHS-Si
stabiliser, which adsorbs directly onto the glass
\cite{ABSPrivComm}.

\subsection{Achieving Optimum Images \label{imagecapture}}
In this section we discuss how to achieve images of bulk
(three-dimensional) samples in a digital format which are suitable
for particle location. Notably, we discuss how to choose an
appropriate pixel pitch to best balance fidelity and image size.  We
outline a recipe for capturing good quality images, and some common
sources of poor image quality.

\subsection*{Contrast}
The intensity of light is usually detected by photomultiplier tubes
(PMTs) or charge coupled devices (CCDs). After analog-digital
conversion, the image is represented by a series of discrete levels,
greyscales or greylevels, and the number of greylevels is the image
depth. Visually pleasing images require that gradations between
neighbouring greylevels are barely perceptible; this needs of order
$100$ greylevels. Frequently $256$ greylevels, i.e. $8$-bit images,
are chosen, but also $10$-, $12$-, and $16$-bit images are used.
These larger image depths reflect that `visually pleasing' is not
always sufficient for quantitative studies.

The recorded greylevel is determined by the illuminance (often
loosely the power), which is device specific (e.g. the laser power
entering the confocal microscope), as well as the gain, which scales
the detector output signal, and the offset, which is an additive
constant to compensate for a background count. The latter two are
usually user-definable and should be optimised to maximise image
contrast. Contrast is variously defined, but always describes the
range of intensities present in an image. One definition of the
contrast, $C(x,y,z)$, is:
\[C(x,y,z)=\frac{I(x,y,z)-I_0}{I_0},\]
where $I_0$ is the image mean background intensity, and $I(x,y,z)$
is the intensity at point $(x,y,z)$.

\begin{figure}
\begin{center}
\includegraphics[width=12cm,bb=0 0 500 522]{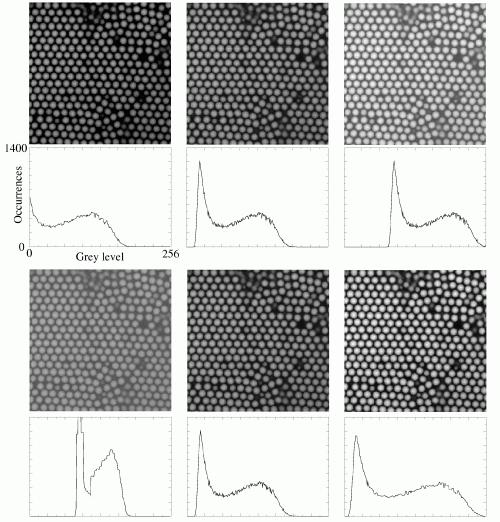}
\caption{Images and corresponding histograms to illustrate the
effect of offset (top) and gain (bottom).  \mod{The sample contains
NBD-dyed PMMA particles with a radius of $1.08\mu m$ in a mixture of
cis-decalin and CHB at a volume fraction of $\phi=0.545$.
Tetrabutylammonium chloride has been added to screen the slight
charge of the particles \cite{mythesis}. The field of view is about
$41{\times}41\mu m^2$ ($256{\times}256$ pixels at $0.16\mu m$ per
pixel), and the original image was captured at a rate of about $0.6$
fps.}\label{histograms1}}
\end{center}
\end{figure}

\begin{figure}[h]
\begin{center}
\includegraphics[width=10cm,bb=0 0 800 384]{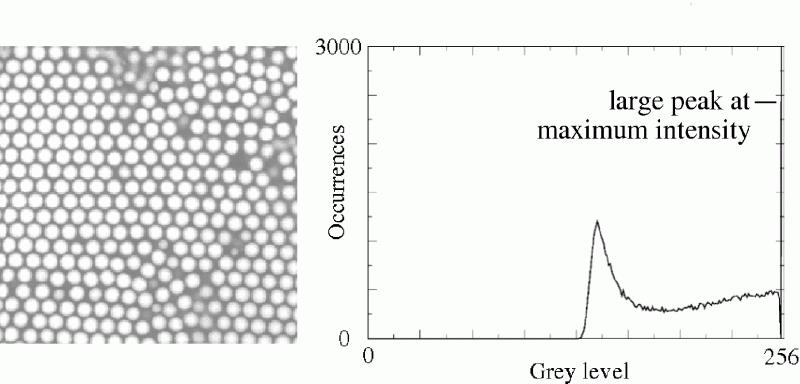}
\caption{Image suffering from saturation, and corresponding
histogram. \label{histograms2}}
\end{center}
\end{figure}

Maximum contrast means using the full dynamic range of the imaging
system.  This is facilitated by using an image histogram, or
occurrence count, of intensity values. Figure \ref{histograms1}
shows typical images and their histograms, which here have two peaks
representing particles and solvent, respectively.  The histogram for
the original image (top and bottom middle) shows an appropriate
range of brightness, while the top and bottom rows illustrate the
effect of altering the offset and gain, respectively. \mod{While the
original image (top and bottom middle) shows an acceptable
histogram, that on the bottom right uses the full dynamic range of
the detector best and is thus closest to what one should aim for.}
 Furthermore, one must avoid saturation, where the highest intensity
pixels ought to take a value greater than the image format maximum
($255$ for $8$-bit images) and therefore are artificially restrained
to this maximum value (Figure \ref{histograms2}).

Note that the operations shown in Figures \ref{histograms1} and
\ref{histograms2} were performed after image capture and thus merely
simulate the effect of adjusting the gain and offset; they are
lookup table (LUT) operations. This raises an important point: LUT
operations do not increase the information contained within the
images.  These are empty operations, and serve only to increase the
visual appeal of the images.  To maximise the information retained
from the imaged volume the imaging system parameters must be set so
that the contrast in the detected image corresponds to the dynamic
range at the time of capture.

\subsection*{Pixel Pitch and Image Size\label{pixelpitch}}
As well as ensuring the image uses the full dynamic range of the
detector, one must choose an appropriate pixel pitch and region of
the sample to study. Recalling the Nyquist-Shannon requirement
(\S\ref{Nyquist}), the pixel pitch should be around $100$ --
$150$~nm laterally and $250$ -- $300$~nm axially. The particles are
isotropic, however, and it is convenient to have the pixel pitch
close to isotropic. Moreover, the particle location schemes we will
discuss below are suited to particle sizes of approximately $11$ --
$13$ pixels in each direction. For particles of diameter $2~\mu$m,
this suggests a pixel pitch of $150$ -- $180$~nm in each direction.
Since the particle shape is known, particle locations can be
inferred to greater resolution than the sampling frequency suggests,
so that a slightly larger pixel pitch can be chosen.

For a quantitative analysis, a reliable means of calibration of the
pixel pitch is essential. Standard test beads (e.g. TetraSpeck
Fluorescent Microspheres) allow lateral and axial calibration
and identification of spherical aberrations, but are usually too
small for a reliable calibration; for a $4~\mu$m diameter particle
and a pixel pitch of $0.15~\mu$m the error in the pixel pitch is
around $2\%$. A high resolution test slide (e.g. Richardson
Test Slide, Model 80303) is better suited for two-dimensional
calibration as well as identification of distortions; distances of
order $10~\mu$m can easily be calibrated, determining the pixel
pitch to better than $1\%$. Using this method, it is also possible
to establish variations in the pixel pitch across the images.

Having determined the pixel pitch, one must decide upon the desired
field of view.  Unlike in conventional microscopes, in scanning
microscopes the size of the region scanned, and thus the field of
view, can be user-defined. The pixel pitch and field of view then
determine the number of pixels in the image, which in turn is
limited by the image processing hardware.  (With our current desktop
computing hardware, the limit is typically around
$512{\times}512{\times}100$ pixels, for $16$ bit images, with this
requirement being relaxed as desktop computing performance,
particularly memory size, improves.) The ideal choice of parameters
is therefore a compromise between the sampling requirement, the
desire for about $11$ -- $13$ pixels per particle in each direction
and the overall image size (or number of pixels). For the above
mentioned system a voxel size of around
$0.13~\mu$m${\times}0.13~\mu$m${\times}0.2~\mu$m is convenient and
suitable, and results for $512{\times}512{\times}100$ voxels in a
visible volume of about $66~\mu$m${\times}66~\mu$m${\times}20~\mu$m,
which is large enough for many purposes.

\subsection*{Capturing Images\label{goodimagerecipe}}
Initially bright-field illumination is useful to find a suitable
focus in the sample, nearest to the cover glass is most
straightforward. Bright-field illumination affords a greater field
of view, which, together with the lack of confocal sectioning, makes
it much easier to `find' the particles. Changing to confocal imaging
should immediately give an image, although most likely a poor one.

With depth in the sample, scattering decreases the signal and hence
the SNR. Images captured deeper in the sample are thus inherently
more noisy than shallower ones.  Data should therefore be captured
as close to the cover slip \mod{as the phenomenon under observation
will allow (or indeed requires)}.  To avoid saturation, the imaging
parameters, such as laser power, gain and offset, should be set at
the shallowest (i.e. brightest) point of the region to be imaged.
Though a nuisance, we can offset the effect of scattering against
that of photobleaching. By starting a scan deep in the sample, where
the image is of relatively low intensity but not yet photobleached,
and proceeding to a shallower region, any photobleaching will tend
to counteract the increase in the intensity. (Though the excitation
light is reasonably well concentrated in the focal volume,
photobleaching is not confined to the focal plane.) For this reason,
when using an inverted microscope we use a coordinate system with
the positive direction pointing `down' with respect to the
laboratory.

Among the adjustments in the confocal microscope, the size of the
confocal aperture deserves special attention.  Smaller apertures
\mod{result in a smaller depth of field and thus} give `more
confocal' images, while larger ones provide more collected light.
Particularly when the shape of the particles is known the gain in
light collected and therefore SNR may offset the loss of resolution
due to the broadened PSF. \mod{The amount of collected light also
depends on the acquisition time, which should be significantly
shorter than any relevant time scale of the sample \cite{Prasad07}.}

\section{Finding Particle Coordinates \label{findingcoords}}
\subsection{Dealing with Noise\label{noisedealing}}

Here we describe how to deal with noise prior to feature extraction.
Apart from slight modifications, this procedure follows a method
described previously \cite{Crocker96}. Noise may include geometric
distortions, non-uniform contrast and instrumental noise. We will
not discuss geometric distortions, the absence of which can be confirmed
using a test slide; supposing they were present, they can be
dealt with using standard algorithms \cite{Crocker96}.

\subsection*{Contrast Gradients}
The procedure we describe was originally intended to process images
captured using CCD cameras.  Such images, in which different pixels
are sampled by different detector elements of, in general, different
sensitivity, frequently display contrast variations. Even if in most
confocal microscopes all of the pixels are scanned by the same
detector, this procedure corrects for non-uniformities in illumination
across the field of view.

Provided the image contains features which are suitably small and
sufficiently far apart, large scale variations in the background (we
take this to mean on a length larger than the particles) can be
adequately modelled by a `boxcar' average of extent $2w{+}1$, where
$w$ is an integer larger than the sphere's apparent radius in
pixels, but less than the typical intersphere separation and can be
different in the three dimensions \cite{Crocker96}. This corresponds
to a real-space convolution of the image with the following kernel:
\begin{equation}
A_w(x,y,z)=\frac{1}{(2w+1)^3} \sum_{i,j,k=-w}^w A(x+i,y+j,z+k)
\label{bkernel}
\end{equation}

This correction relies on the assumption that features are `small'
and `well separated', which, loosely, means that the typical
intersphere separation is larger than the sphere. For samples
of volume fraction $\phi {\simeq} 0.5$ or greater, this is not true.
Nonetheless, we see later that the results are reasonable.

We note that contrast gradients may genuinely be present.  For
example, in an image of a crystalline sample, crystallites may lie
in slightly different planes from one another, giving rise to a
genuine contrast variation. We assume that such effects are
negligible.

\subsection*{Single Pixel Noise\label{dealingwithnoise}}
For confocal micrographs of colloidal samples, the instrumental
noise often comprises a significant proportion of single pixel
noise. This would ordinarily be dealt with using a median filter
\cite{GonzalezWoods}; each pixel is replaced with the median value
from its neighbourhood, whose size is typically chosen to be $3$,
$5$, or $7$ pixels. This has a smoothing effect which deals well
with single pixel noise.  In keeping with many earlier studies, we
prefer another approach. It is difficult to defend this over the
median approach {\em a priori}, but it turns out to work well, and
is remarkably robust.

We assume single pixel instrumental noise, or, equivalently, that
the noise has correlation length $\lambda_n {\simeq} 1$ pixel.
Removal of all features having this lengthscale by low pass
filtering would certainly eliminate single pixel noise, but this has
the disadvantage of blurring edges\footnote{Low-pass filtering, in
its simplest form, involves cutting from the Fourier Transform of
the image all points that lie above a threshold frequency.  Such a
circular region in the Fourier Transform gives rise, upon Fourier
transforming once more, to a real-space convolution kernel of the
form $A_{\mathrm{lowpass}}{=}J_1(r/w_0)/(r/w_0)$, where
$r^2{=}x^2{+}y^2{+}z^2$ and $w_0$ is the threshold frequency. This
kernel effectively places a $jinc$ function, or series of concentric
rings, about each point in the original image, thereby blurring
beyond usefulness the processed image.}.  Rather, the usual approach
is convolution of the image with the kernel:
\begin{equation}
A_{\lambda_n}(x,y,z)=\frac{1}{B} \sum_{i,j,k=-w}^w
A(x{+}i,y{+}j,z{+}k)\:
exp\left(-\frac{i^2{+}j^2{+}k^2}{4\lambda_n^2} \right)
\label{gkernel}
\end{equation}
where $B$ is the normalisation condition, $B =\left[
\sum_{i=-w}^w exp\left(-(i^2/4\lambda_n^2) \right) \right]^3$. Since
the Fourier Transform of a Gaussian is itself a Gaussian, this
attenuates high frequencies as desired, while more adequately
preserving edges.

It was shown \cite{Crocker96} that that these two operations (equations
\ref{bkernel} and \ref{gkernel}) can be implemented in a single step
using the following kernel:
\begin{equation}
K(i,j,k)=\frac{1}{K_0}\left[ \frac{1}{B}
exp\left(-\frac{i^2+j^2+k^2}{4\lambda_n^2} \right)
-\frac{1}{(2w+1)^3} \right]
\end{equation}
The normalisation $K_0 = 1/B\left[ \sum_{i=-w}^w
exp\left(-(i^2/2\lambda_n^2) \right) \right]^3-(B/(2w{+}1)^3)$ is
appropriate for comparison between images filtered with different
values of $w$.  $\lambda_n$ is again set to unity.

A practical limitation on the convolution arises since it involves,
for each point $(x,y,z)$ in the original image, a sum over all
points $(x{+}i,y{+}j,z{+}k)$ for $i,j,k=-w\dots w$.  This cannot
occur for any point which lies less than $w$ pixels from the edge of
the original image.  In practice these are often discarded.  Depending on the application,
it may be useful to have coordinates from this region despite the reduced information.
To achieve this, the original
image is `padded out' with a border of width $w$ around the entire
volume. (Above and below the stack the pixels are all set to the
mean intensity value in the first and last slices respectively. To
the sides of the stack, the border around each slice is set to the
average value of the intensity in that slice.) This padding permits
the entire image to be retained, while the padding is discarded
afterwards.  This average is necessarily not ideal and somewhat arbitrary (one cannot
generate information which is not in the image initially), making
particle coordinates within the border inherently less reliable than
those from the bulk sample.  Interestingly, failure to carry out the padding introduces artefacts
into the Fourier Transforms of the images; if the Fourier
representation is useful, this may be important.  For details of the
artefact, see \cite{mythesis}.

\begin{figure}[h]
\begin{center}
\includegraphics[width=10cm,bb=0 0 600 304]{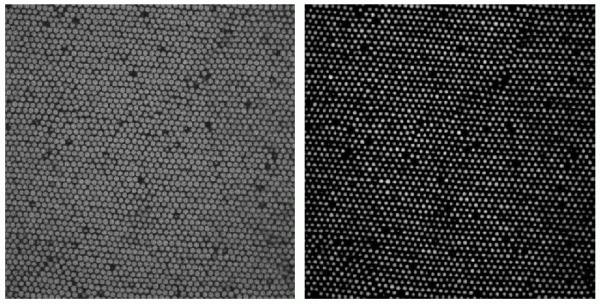}
\caption{A good quality but noisy image before (left) and after
(right) processing. Large-scale contrast gradients are not evident
in this case.  \mod{The particle size was around $10$ pixels, and
the corresponding kernel size was $w=[11,11,9]$, and
$\lambda_n=[1,1,1]$; for details of the parameter choice, see
section \ref{paramopt}.}\label{origimage}}
\end{center}
\end{figure}

Figure \ref{origimage} (left) illustrates a slice from a typical
good quality stack of a largely crystalline region.  The volume was
then filtered using the three-dimensional algorithm, and the same
slice extracted once more (Fig.~\ref{origimage} right). This example
is of a successfully filtered image.  In dense samples, neighbouring
particles occasionally show `bridges', or small bright bands
connecting their images.  This is an important but unavoidable
problem which we discuss in detail later.

\subsection{Strategies for Finding Particle Coordinates}
The centre of a particle can only be identified from its image by
relying on {\em a priori} knowledge of its shape. In our case each
particle is spherical, and, following the arguments above, we thus
know the shape of the fluorescence intensity profile through each
particle's image. Use of {\em a priori} knowledge in this way is
known as Bayesian inference, and is widely used in many fields
\cite{DAgostini03,Dose03}. Even the simplest particle location
scheme infers coordinates by taking advantage of this information
and is therefore in this sense Bayesian.

{\em A priori} knowledge of the particles' shapes permits
location of the particles to higher precision than the sampling
rate.  This sub-pixel resolution ultimately allows location of
particles to substantially better than the resolution of the
microscope. In principle there is no limit on this statement, though inherent
experimental error and limitations in the interpolation techniques
ensure a maximum resolution which cannot be known in advance;
we address this later.

We consider three strategies to identifying particle locations.  The
first involves identifying particles and subsequently inspecting
each in turn (Sec.~\ref{findbrightsandrefine}). This is widely used,
but can only be useful when finding the centres of
solidly-fluorescent spheres. The second, a deconvolution method to
extract instances of the particle image
(Sec.~\ref{SSFdeconvolsection}), is more general, but not used here.
Similarly, the third, the Hough Transform (Sec. \ref{HoughTransform}), is
very general and, we argue, suitable for further exploitation in
colloidal studies, but not used here.  A recent study carried out a
computational comparison of some of the methods described below
\cite{Cheezum01}.  They found, perhaps unsurprisingly, that a direct
Gaussian fit to the intensity profile was most successful for point
sources (where the measured intensity profile ought to be simply the
PSF), whereas a pattern-matching approach (see `Refinement using
Measured SSF', below) was more successful for larger particles.
They did, however, neither consider real data nor the more
complicated situation where several particles are present.

\subsubsection{Identification of Local Brightness Maxima and Subsequent Refinement \label{findbrightsandrefine}}
The majority of particle location schemes operate on the assumption
that the image of a particle has a maximum intensity at, or near to,
its centre, i.e. that there is a one-to-one correspondence between
local brightness maxima and particle centres.  (`Local' is important
since the image brightness can in general vary dramatically on the
scale of several particle diameters without compromising the technique.)
In practice, however, the sampling grid will never coincide with the
sphere centre and also other imperfections, for example resulting
form noise, have to be considered. Nevertheless, we assume that
based on the local brightness maxima we can reach nearest
pixel accuracy.  The refinement step then gives subpixel resolution.
There is a hierarchy of possible refinements, based on the extent
to which the {\em a priori} knowledge is relied upon.

\subsubsection*{Refinement Using the Spherical Symmetry of the SSF: Centroiding}
The simplest technique relies only on the knowledge that the SSF is
spherically symmetric. This is by far the most widely-used
technique \cite{Raffel98} and will be described in detail in
Sec.~\ref{centroiding}.

\subsubsection*{Refinement Using a Functional Form for the SSF \label{threepointestimators}}
Rather than using a centroiding approach, i.e. the fact that the SSF is spherically
symmetric, the functional form of the SSF could be used. Already an
approximation to the functional form might improve the centroiding
approach. In particular, it improves the performance in the case
of SSF overlap, which is modeled by a simple superposition.

In related systems, such as particle image velocimetry (PIV), as well
as the $n$-point centroid estimators, there are two widely used
approximations to the functional form \cite{Raffel98}.  The first is a
parabolic peak fit with the functional form
\begin{displaymath}
s_{par}=Ax^2+Bx+C,
\end{displaymath}
and similarly for $y$ and $z$.
Based on this model, the `true' position ($x_0,y_0,z_0$) of the particle is:
\begin{eqnarray*}
x_0&=&i+\frac{f_{(i-1,j,k)}-f_{(i+1,j,k)}}{2f_{(i-1,j,k)}-4f_{(i,j,k)}+2f_{(i+1,j,k)}}\\
y_0&=&j+\frac{f_{(i,j-1,k)}-f_{(i,j+1,k)}}{2f_{(i,j-1,k)}-4f_{(i,j,k)}+2f_{(i,j+1,k)}}\\
z_0&=&k+\frac{f_{(i,j,k-1)}-f_{(i,j,k+1)}}{2f_{(i,j,k-1)}-4f_{(i,j,k)}+2f_{(i,j,k+1)}}.\\
\end{eqnarray*}
where $(i,j,k)$ is the candidate (integer) location and
$f(i',j',k')$ is the intensity of the sampled image at position
$(i',j',k')$. In our work, there is no basis for using a parabolic
fit.

A Gaussian fit seems more appropriate \cite{Crocker96}, although the
deviations are sometimes significant (Sec.
\ref{comparisonSSFmodelledreal}, \cite{Baumgartl05}):
\begin{displaymath}
s_{Gauss}=C\exp{\left[ \frac{-(x_0{-}x)^2}{k} \right]}.
\end{displaymath}
This implies that
\begin{displaymath}
\ln{s_{Gauss}}\propto-(x_0{-}x)^2,
\end{displaymath}
so that the three-point Gaussian estimate is parabolic in the
natural logarithm of the sampled points $f(i,j,k)$:
\begin{eqnarray*}
x_0&=&i+\frac{\ln f_{(i-1,j,k)}-\ln f_{(i+1,j,k)}}{2\ln
f_{(i-1,j,k)}-4\ln f_{(i,j,k)}+2\ln f_{(i+1,j,k)}}
\end{eqnarray*}
and similarly for $y_0$ and $z_0$. These three-point estimators are
convenient and widely applied. They rely on the image of the
particle being around three pixels in diameter \cite{Raffel98},
which may not permit the desired sampling rate: remembering Section
\ref{pixelpitch}, the pixel pitch ought to be around $0.2~\mu m$, so
a typical colloidal particle suitable for confocal microscopy
(diameter ${\simeq} 2~\mu m$) need be at least $10$ pixels in
diameter.  Once again, knowledge of the functional form of the SSF
may allow recovery of the particle location despite the apparent
undersampling.  It is not possible in principle to resolve this
conflict, and opinions, judging by the literature
\cite{Royall03,Campbell02}, differ on exactly what is the
appropriate choice.

Fitting to a functional form outperforms the centroid approach
when the SSFs of neighbouring particles overlap.  It is well
documented that the above fits are appropriate only for well-resolved correlation
peaks \cite{Raffel98,Bolinder99}.  Progress in this direction
has not been attempted, largely because the
functional form is not in general known. An approach which avoids
this deficiency is described next.

\subsubsection*{Refinement Using the Measured SSF}
In this case, the SSF is measured in a window just larger than
the SSF itself.  `Stamps' of this SSF around the candidate
particle locations are then fitted to the image. Provided the
SSF was sampled appropriately, it also accounts for
aberrations and imperfections, and more
accurately for the system PSF. We describe this technique in
Section \ref{SSFrefinement}.

\subsubsection{Particle Location by Deconvolution of the
SSF\label{SSFdeconvolsection}}

If we are to extract instances of a motif, as the above refinement
using the measured SSF suggests, then a deconvolution technique is
seemingly more appropriate. We have established that the imaging
system can be represented as a convolution process
(Sec.~\ref{ImagingProcess}, Eq.~\ref{eqPSF}), and we could in
principle recover the original form of an imaged object by
deconvolution of the PSF from the observed image. Similarly, the
image of a particle is formed by placing a copy of the SSF (the
above `stamp') at the particle coordinate
(Sec.~\ref{comparisonSSFmodelledreal}, Eq.~\ref{eqSSF}). From
Equation \ref{eqnSSFconv} it follows that
\begin{displaymath}
\delta(x,y,z)=\mathcal{F}^{-1}\left\{\frac{\hat{F}}{\hat{S}}\right\},
\end{displaymath}
where $\mathcal{F}^{-1}\{\}$ denotes the reverse Fourier Transform.
Thus by deconvolving the SSF from the image of a particle, we obtain
a single bright point at its centre which is easily located.

If there are several spheres in the image, the situation is
complicated slightly, since it is not possible to define the centres
of all spheres as being at the origin.  The Fourier Transform of a
delta function which is not centred on the origin is:
\begin{displaymath}
\mathcal{F}\{\delta(\mathbf{r}{-}\mathbf{r}_0)\}(\mathbf{k})=\int^{\infty}_{\mathrm{r}=-\infty}\delta(\mathbf{r}{-}\mathbf{r}_0)e^{-2\pi
i\mathbf{k}. \mathbf{r}}\mathrm{d\mathbf{r}}=e^{-2\pi
i\mathbf{k}.\mathbf{r}_0}.
\end{displaymath}
Thus the Fourier Transform is no longer constant, but contains phase
information which encodes each particle's distance from the origin.
This is the reason why the locations of several particles can be
determined by the deconvolution process.

In the case where the field of view contains $N$ particles at
positions $\mathbf{r}_i$, the image can be written
(Eq.~\ref{eqSSF}):
\begin{displaymath}
f(x',y',z')=\sum_{i=1}^{N}\delta(\mathbf{r}{-}\mathbf{r}_i)\odot
s(x,y,z)\;.
\end{displaymath}
Since the Fourier Transform is a linear operation, this gives us
(Eq.~\ref{eqnSSFconv})
\begin{displaymath}
\hat{F}(u,v,w)=\alpha_1\hat{S}+\alpha_2\hat{S}+... = \hat{S}
\sum_{i=1}^{N}\alpha_i,
\end{displaymath}
with $\alpha_n$ the Fourier transform of the $n^{\mathrm{th}}$
delta function. 
Finally we obtain
\begin{equation}\label{SSFdeconvol}
\mathcal{F}^{-1} \left\{ \sum_{i=1}^{N}\alpha_i
\right\}=\mathcal{F}^{-1}
\left\{\frac{\hat{F}}{\hat{S}}\right\}=\sum_{i=1}^{N}\delta(\mathbf{r}{-}\mathbf{r}_i).
\end{equation}
Thus by measuring the sphere spread function $s$ carefully, we could
in principle deconvolve it from the observed image $f$, to obtain a
series of bright points indicating the positions of the imaged
spheres.  Locating the bright points then returns the positions of
the particles.

Although this technique is appealing, it is very sensitive to noise in
the detected image, similar to the deconvolution of the PSF \cite{mythesis}.
The effect of noise in Fourier space is (Eq.~\ref{eqNoise}):
\[\hat{G}(u,v,w)=\hat{F}(u,v,w)+\hat{N}(u,v,w).\]
Noise is usually additive and highly localised in the object space,
i.e. single pixel.  In Fourier space, the noise is therefore highly
delocalised, and its amplitude $|\hat{N}(u,v,w)|^2$ nearly constant.
Dropping arguments we write
\[
\frac{\hat{G}}{\hat{S}}=\frac{\hat{F}}{\hat{S}}+\frac{\hat{N}}{\hat{S}}
                       =\frac{\hat{O}\hat{S}}{\hat{S}}+\frac{\hat{N}}{\hat{S}} =
\hat{O}+\frac{\hat{N}}{\hat{S}}.
\]
If there is no noise present, we recover the expected form
(Eq.~\ref{SSFdeconvol}).

This technique shares a problem inherent in many deconvolution
processes: Wherever the function $\hat{S}$ falls near to
zero, the first term on the right hand side is liable to become very
large (depending on the behaviour of $\hat{F}$ at that point), while
the second term is certain to become very large, since
$\hat{N}$ is approximately constant.  Because many deconvolution
kernels contain zero-height pixels, deconvolution of a noisy image
cannot be relied upon in this simple implementation.

There are several schemes for circumventing this difficulty. For
example, one can seek the solution which contains the same
information in Fourier and real space (Weiner or Optimal Filter).
Furthermore, Maximum Entropy techniques determine as smooth an image
as is consistent with the original data. To our knowledge, these
techniques have been sparingly used in colloidal systems
\cite{Brujic03,Brujic03b} and we thus refer to the literature for
details \cite{GonzalezWoods}.

\subsubsection{Hough Transform\label{HoughTransform}} The Hough
Transform is a feature extraction technique popular in computer
vision \cite{GonzalezWoods}. The original version identified lines
\cite{Hough62}, and this has been generalised to find the outlines
of arbitrary shapes \cite{Duda72,Ballard81}. It can be applied to
differently shaped objects and is in particular not restricted to
particles with a bright centre. However, at least in its simple
form, a suitable parameterisation for the outline of the shape has
to be found. (In the generalised version, the parameterisation is in
the form of a lookup table.)  Briefly, the Hough Transform operates
by inverting the parameterisation so that the parameters become the
coordinates (in Hough space) and the real-space coordinates become
parameters.  When points in real space are transferred to Hough
space, any regions of high point density indicate objects and their
parameters.  This is assessed using accumulator cells, or bins,
whose size dictates the precision to which the parameters are found.

The Hough transform has been applied successfully to systems of
discs \cite{Warr96,Warr94} (in fact, spheres viewed in
two-dimensions), and would presumably translate well to colloidal
studies, but we are not aware of any three-dimensional studies. This
might be due to the difficulty of extending the Hough Transformation
to systems with many parameters, in particular the need for many
points per accumulator cell for reliable parameter determination. In
addition, the Hough Transform is dependent on the object outlines
(edges) having been detected reliably.

\section{Tests of Accuracy\label{accuracychecks}}
Before we present the centroiding and SSF refinement techniques in
detail (Sec. \ref{centroiding} and \ref{SSFrefinement}), we review
some tests which can be used to compare the accuracy of both
techniques.  At least the first three of these are well known
\cite{Crocker96,WeeksTrackingWeb}.

\subsection{Basic Tests}
\begin{figure}[h]
\begin{center}
\mbox{
\includegraphics[width=2.5cm,angle=-90,bb=0 0 265
473]{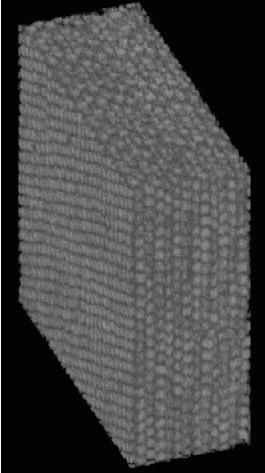}\qquad
\includegraphics[width=2.5cm,angle=-90,bb=0 0 264 357]{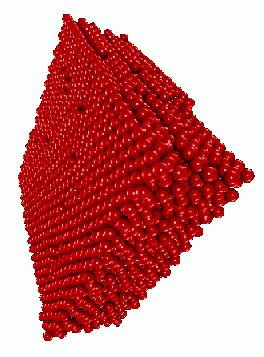}}
\caption[An example of a reconstruction based on particle
coordinates.]{Confocal micrograph (left) and corresponding
reconstruction (right) based on particle
locations.\label{samplecrystal}}
\end{center}
\end{figure}

There are two basic visual tests.  Firstly, a reconstruction that
places an image of a sphere at its supposed position gives a
very crude indication of believability
(Figure \ref{samplecrystal}), although this gives no
objective measure of accuracy.

\begin{figure}[h]
\begin{center}
\includegraphics[width=10cm,bb=0 0 600 599]{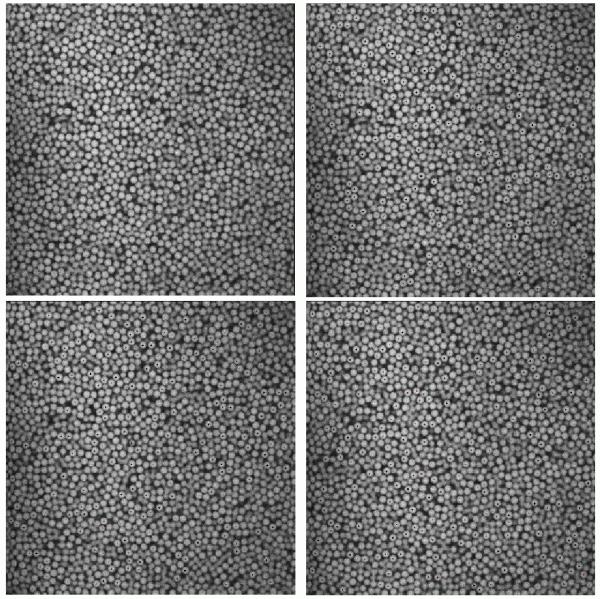}
\caption[A typical two-dimensional slice from a sediment with
crosses overlaid at the particle locations.]{Two-dimensional slice
(top left) and the same image with crosses placed at the (nearest
pixel to the) detected location. Particles seemingly missed are
detected in adjacent slices in the sequence
(bottom).\label{samplefover}}
\end{center}
\end{figure}

The one-to-one correspondence between particle images and detected
locations can be checked using markers (e.g. crosses) on the
original image (Fig.~\ref{samplefover}). Without an empty
magnification it is, however, only possible to overprint markers to
the nearest pixel.


\subsection{Test for Rogue Particles}
\begin{figure}[h]
\begin{center}
\includegraphics[width=12cm,angle=00,bb=0 0 800 260]{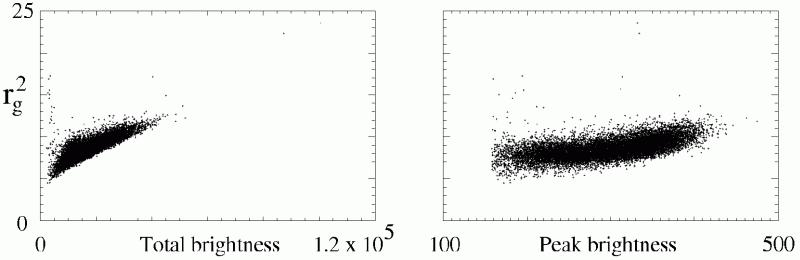}\quad
\caption[Two example clouds showing radius of gyration squared, and
peak and integrated brightness.]{Properties characterising particle
images: radius of gyration squared of a particle versus its total
brightness (left) and its peak brightness (right). \label{clouds}}
\end{center}
\end{figure}

For a perfectly imaged sample, each particle image should
be identical; variations may indicate imperfectly located particles.
Simple properties characterising the particle image are its
total brightness, peak brightness, the first moment of its
intensity distribution (its radius of gyration), and its
ellipticity \cite{Crocker96}. For identical particles, we
expect these values to fall within a narrow band of
values. Figure~\ref{clouds} shows two examples, the
radius of gyration squared of a particle versus its total
brightness (left) and its peak brightness (right). Most
points are close together within a tight locus
except a few, which might be disregarded as rogue
particles. However,
it is never certain what the acceptable locus is, and
correspondingly never obvious where the cutoff should occur.
Furthermore, this does not allow us to infer the reliability of
particle locations on an individual basis.


\subsection{Test for Pixel Biasing}
\begin{figure}[h]
\begin{center}
\includegraphics[width=14cm,angle=0,bb=0 0 800 214]{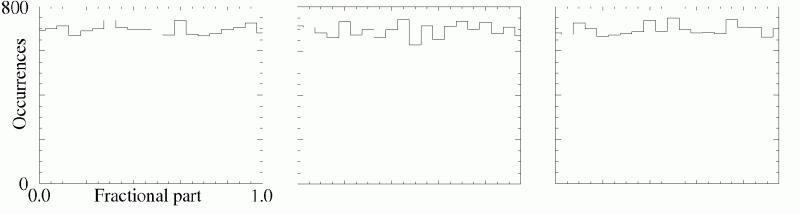}
\caption[Histograms of the fractional part of the coordinates in x,
y, and z.]{Distribution of the fractional part of the particle
locations found using the centroiding technique.
\label{pixelbiasing}}
\end{center}
\end{figure}

All of the techniques we consider here involve finding the nearest
pixel to the true centre first and then refining this candidate
particle location. If the first step fails, particle locations are
not reliable, which can be detected, for example, by the above
means. However, if the first step is successful but the refinement
step fails, the particle locations are sometimes biased towards
integer values. We can detect this using the distribution of the
fractional part of the particle location, which should be evenly
distributed from zero to one (Fig.~\ref{pixelbiasing}).


\subsection{Tests based on structural properties\label{structproperties}}
Often it is possible to compare the structural properties of an
ensemble of particles with the expectation either from theoretical
studies, or from simulations.  Although there are many possible
structural descriptors, the radial distribution function, $g(r)$, is
most commonly used to check the reliability of the particle
location. The height of the first peak and its sharpness, i.e. where
it begins to rise (which ought to be infinite and $2r$ respectively
for hard spheres), are useful indicators of the reliability of the
determined particle locations. Nevertheless, is it rather a measure
of location precision than accuracy and it is also an average
quantity, and hence unable to test individual particle locations. We
will use this criterion in Secs.~\ref{centroiding}
and~\ref{SSFrefinement}.

\subsection{Checking the volume fraction \label{localvolume}}
\mod{The volume fraction of a sample as prepared is often only known
with a large systematic error \cite{mythesis}.  It can therefore be
helpful to calculate it from the determined particle coordinates,
which can also serve as a consistency test for the coordinates.}

The total volume fraction $\phi$ can be determined from the volume
of the imaged region, $V_{box}$, the number of particles it
contains, $N$ as well as their radii, $R$:
\[\phi = \frac{4}{3}\pi R^3
\frac{N}{V_{box}}.\] This depends on a reliable determination of the
particle number $N$, i.e. that neither particles are missed nor
spurious particles are added. To overcome this problem in spatially
homogeneous samples, we can determine the local volume fraction for
each particle. This requires us to determine the volume per
particle, which can be found by partitioning the space
appropriately; there is no unique way of doing this, but the
Vorono\"{\i} construction \cite{PreparataBook} is well-defined and
physically sensible \cite{mythesis}.  For a homogeneous sample the
local volume fraction ought to be similar (within a certain
distribution) for all of the particles.  Anomalously large volumes
will be attributed to neighbours of a missed particle, which can be
ignored in calculating the mean volume fraction.  The volume per
particle, and its distribution, may itself be of interest in
spatially heterogeneous samples.

 \mod{The value of $\phi$ as determined from the particle coordinates is subject to a
systematic error, due to the error in the determination of the
particle radius, $R$. Typically, an error of around one percent in
$R$ can be achieved (correspondingly, three percent in $\phi$). It
is usual to get the random error in the coordinate-derived $\phi$ to
much better than this.  For details of a comprehensive dataset, see
\cite[Figure 8.1]{mythesis}.}

\section{Centroiding\label{centroiding}}
In this section we describe centroiding as commonly applied to
colloidal systems.   We note that centroiding methods are used in
several other fields. Most notable is particle image velocimetry
(PIV) \cite{Raffel98}, in which roughly-Gaussian shaped
(correlation) peaks are located
\cite{Elkins77,Racca88,Willert91,Westerweel93,TedSchlicke02} with
some using more sophisticated procedures, such as the three-point
estimator (Sec.~\ref{threepointestimators})
\cite{Raffel98,TedSchlicke02} or iterative procedures
\cite{Sugii00}. They also include discussions on the accuracy of
centroiding \cite{Bolinder99} as well as the presence of noise
\cite{Cao94,Thomas,Ares04}. The centroiding method was first applied
to two-dimensional projections of colloidal samples by Murray and
Grier \cite{Murray96}, and then refined in the classical paper by
Crocker and Grier \cite{Crocker96}, which also forms the basis for
the routines by Weeks \cite{WeeksTrackingWeb}. All centroiding
algorithms seek to find with maximum accuracy the position of the
centroid of an image distribution, but in colloidal studies this is,
as we will explain, not always what we desire.  Although this
section is particularly relevant to Weeks' tracking routines, it is
not uniquely so and should also be relevant to most other
centroiding routines.

\subsection{Procedure\label{centroidingprocedure}}
After filtering the raw image (Sec.~\ref{dealingwithnoise}), the
image is searched for local maxima to determine the candidate
particle locations, i.e. the nearest pixel (integer) to the `true'
location.  Local maxima are simply the brightest points within a
three-dimensional region (of size $2\times separation$, see below).
Subsequently a window (of size $extent$) is considered around each
candidate particle location. Within this window, the centroid of the
image intensity distribution is taken, giving the final particle
coordinate.  Other properties of the intensity distribution, such as
total brightness, peak brightness and radius of gyration, can easily
be determined within this window.

To avoid problems due to external noise, most notably the
influence of light from objects other than that under study,
the algorithm ignores all pixels within the above window
whose intensity falls below a certain fraction (quantified by
$threshold$) of the peak height belonging to that particle.
Statistics calculated for a particle hence refer only to pixels
having intensity greater than this fraction of the peak intensity.
In addition to this `local' threshold, it can be useful to
disregard any features whose peak brightness is less than a
certain fraction of the brightest feature of the whole image,
on the grounds that these are spurious.  While it is possible
for the brightness of features to vary substantially within an
image, particularly for dense objects which are poorly
refractive index-matched to the solvent, this is a useful
safeguard and commonly chosen (arbitrarily) to be around $0.7$.

\subsection{Parameter Optimisation\label{paramopt}}
The centroiding technique is in general quite robust to variations
in its parameters, as we will show in this section. Since the
parameter space is huge, we concentrate on its most useful portion
and illustrate the effect of the different parameters using a high
quality image. It is a $512{\times}512{\times}100$ image with a
pixel pitch of $0.16~\mu$m$\times 0.16~\mu$m$\times 0.20~\mu$m and
originates from PMMA particles of radius $r{=}1.09~\mu$m and volume
fraction $\Phi{\simeq}0.64$ in {\em cis}-decalin and was taken using
a BioRad confocal microscope. We use as standard parameters:
$separation{=}[6,6,5]$, $extent{=}[13.0,13.0,11.0]$,
$threshold{=}0.5$ and for noise filtering ($w$, as defined in Sec
\ref{noisedealing}) $[13,13,11]$. The effect of the parameter choice
is judged using the radial distribution function $g(r)$. As argued
in Sec. \ref{structproperties}, the height of the first peak and the
distance where the first peak begins to rise. Although the
importance of individual parameters depends on the exact system
under study and the indicators for reliability, the following
discussion reflects general trends.

\subsubsection*{Noise filtering parameter}
\begin{figure}[h]
\begin{center}
\includegraphics[height=14cm,angle=0,bb=0 0 600 715]{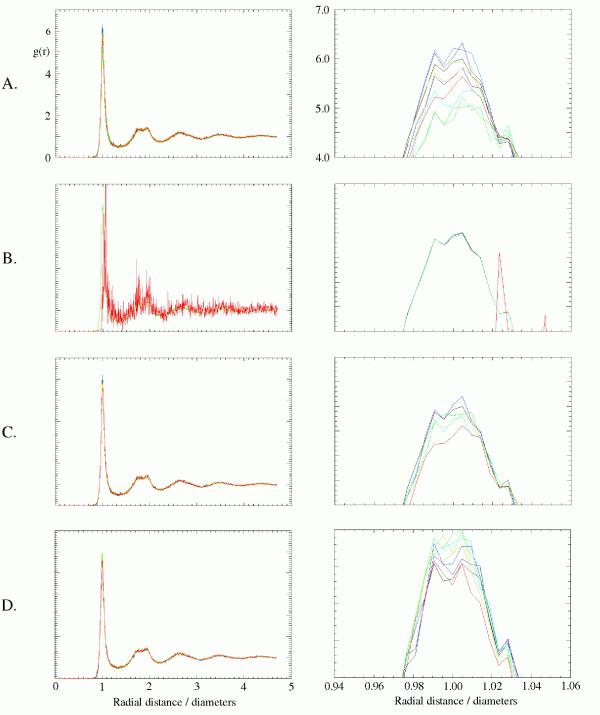}
\caption{Effect of parameters on the radial distribution function as
obtained by determining the particle coordinates using the centroid
method. Varied parameters are (A) noise filter size, (B)
$separation$, (C) $extent$, (D) $threshold$. For values of the
parameters see text. In each case, the left-hand images are the full
rdf, while the right-hand ones are `zoomed in' to a smaller region
to more clearly illustrate the differences.\label{f3dthresh}}
\end{center}
\end{figure}

Figure \ref{f3dthresh}A shows the dependence on $w$, the length
specified in the filtering procedure. Slightly larger windows are
acceptable ($[15,15,13]$, violet, is better than $[13,13,11]$,
black, but very similar to $[17,17,15]$, light blue), while too
small windows should be avoided ($[7,7,5]$ and $[6,6,5]$, both
green). Furthermore, anisotropic windows (above and $[11,11,9]$,
blue; $[9,9,7]$, cyan) are preferable, although isotropic windows
($[13,13,13]$, yellow; $[11,11,11]$, orange; $[9,9,9]$, red) still
work reasonably well.


\subsubsection*{Separation Parameter}
The parameter $separation$ is a $3$-element integer vector.
Typically, the particles are of radius $r{\simeq}1~\mu$m and the
pixel pitch about $0.15~\mu$m laterally and $0.2~\mu$m axially and
thus the particle radius will be around $6$ pixels laterally and $5$
pixels axially. This suggests $separation=[6,6,5]$.

Since this parameter determines (half) the minimum separation
between candidate particle locations and is furthermore used solely
to find candidate particle locations, and is not used in the
refinement step, it will only determine the number of detected
particles. Four cases are shown (Fig.~\ref{f3dthresh}B):
$separation{=}[5,5,4]$ (green), $[6,6,5]$ (black), $[7,7,6]$ (blue)
and $[13,13,11]$ (red), which result in $16417$, $16322$, $16476$
and $2119$ locations respectively. There is only a weak dependence
of the particle number and $g(r)$ on $separation$ provided its value
is near to the particle radius. This is because the particles are
very similar in size; as long as $separation$ is sufficiently small
to capture essentially all of the particles, it has almost no
further effect. $separation{=}[13,13,11]$ is clearly too large and
therefore excludes many bright peaks, $g(r)$ is correspondingly
noisier (as it is based on only about $10$\% of the particles), but
also has its peak shifted upwards by about $5$\%, since the analysis
finds only a subpopulation of apparently larger particles.

\subsubsection*{Extent Parameter}
The window for the refinement should cover a particle and have
an odd integer size. (The latter is for technical, image processing
reasons; a window should have a unique central pixel.) In the
present example we thus choose $extent{=}[13.0,13.0,11.0]$
instead of $[12.0,12.0,10.0]$.

Figure \ref{f3dthresh}C shows that the exact choice of $extent$ is
not critical as long as the window is large enough:
$[13.0,13.0,11.0]$ (black) is very similar to the larger
$[15.0,15.0,13.0]$ (blue), while the smaller $[11.0,11.0,9.0]$
(cyan) is inferior. Also an anisotropic window, $[13.0,13.0,13.0]$
(green), $[11.0,11.0,11.0]$ (yellow) and $[9.0,9.0,9.0]$ (red) is
possible, again provided it is large enough. This is encouraging,
because the size of the window must be chosen by eye from the
original image.

\subsubsection*{Threshold Parameter}
The effect of the threshold parameter depends strongly on the image
quality. In the present example, Fig.~\ref{f3dthresh}D, it is weak.
With increasing $threshold$, the peak height first increases (0.1,
black; 0.2, violet; 0.3, blue; 0.4, cyan; 0.5, green; 0.6, light
green) since pixels which are likely to be spurious are disregarded
(as intended), but then the peak height decreases (0.7, orange; 0.8,
red), which is due to a smaller number of `valid' pixels and thus a
reduced reliability. A compromise value of about $0.5$ is generally
a good choice.  We reiterate that this is a local threshold
parameter used only the refinement step, and unrelated to any global
value such as that described in \ref{centroidingprocedure}.

\subsection{An Appraisal of the Centroiding Technique \label{centroidnotcentre}}
\begin{figure}[h]
\begin{center}
\includegraphics[width=14cm,angle=0,bb=0 0 800 274]{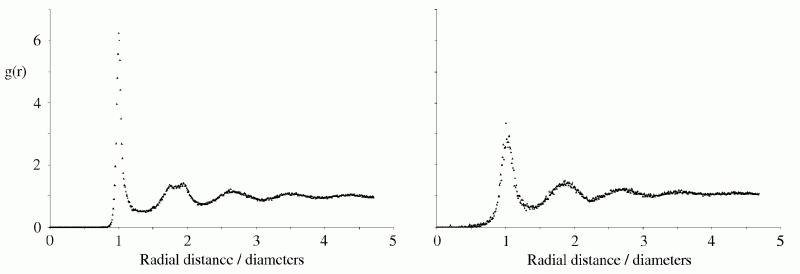}
\caption[Comparison radial distribution functions for `good' and
`bad' glassy samples found using the centroid algorithm.]{Comparison
of the radial distribution functions $g(r)$ for a good quality
(left) and mediocre quality (right) image of two glassy samples of
the same system, as determined using the centroid procedure.
 \label{goodandbadglassyrdf}}
\end{center}
\end{figure}

The centroiding technique is clearly successful, for at least some
sample conditions. For example the radial distribution functions
found from density-matched crystalline samples are convincing
\cite{Campbell02} and also glassy sediments give good results
(Fig.~\ref{goodandbadglassyrdf}, left) even for relatively low
quality images (right), for example those obtained at a high capture
speed.  This figure indicates that these samples give rise to good
$g(r)$ under the right imaging conditions and suggest that poor
quality $g(r)$ may be due to the particle location method rather
than the sample.


However, this technique also has shortcomings.
Usually an accuracy of around $30$~nm in the lateral
and $50$~nm in the axial directions are claimed (for example,
\cite{Weeks00}). Since these values should depend on the
experimental conditions, such as the sample properties, the
noise present and how the image was captured, one needs,
however, to assess the accuracy of each individual image
or, ideally, even each particle location in an image.

The precision with which particle locations have been found can be
inferred by plotting the radial distribution function $g(r)$ with
increasingly smaller bins of radial distance $r$.  Assuming the
sample contains sufficiently many particles, this will result in a
successively sharper peak until the effect of the bin size is no longer
seen. Then the bin size is lower than the mean precision to which the
particle coordinates are known. This precision is typically around
$30$--$40$~nm for the samples used in this enquiry, which is
at best a lower bound on the typical accuracy of particle locations.
However, the precision to which the locations of the centroids of
particle brightness are known is not the same as the accuracy
with which the particle locations are known, because the
centroid of a particle image's brightness is not necessarily the
centre of that particle. This is a consequence of imaging objects
that are close in size to the resolution of the microscope. This
results in images --- or SSFs --- of particles which are larger
than the particles themselves (Fig.~\ref{singleparticle})
and thus the possibility of overlapping images or SSFs.

\begin{figure}[h]
\begin{center}
\includegraphics[width=10cm,bb=0 0 800 231]{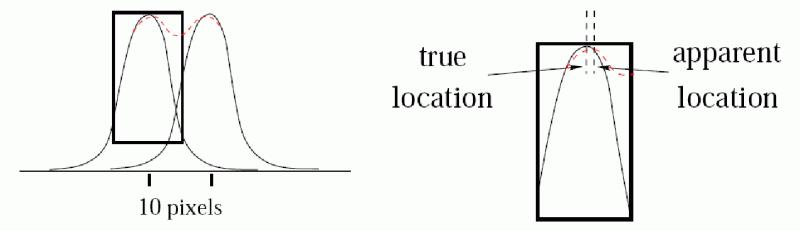}
\caption[The effect of overlapping SSFs on the centroid position.]
{Effect of overlapping SSFs on the centroid position: particle
centres are found too close together.  The right-hand image is a
closer look at the region within the rectangle in the left-hand
image.  The solid black lines represent the individual particle SSFs
and the red dashed line their sum.\label{SSFoverlap}}
\end{center}
\end{figure}

Figure \ref{SSFoverlap} shows the case where two particles are so
close to one another that their SSFs overlap. In this case, the
intensity distribution in the centroid window is clearly asymmetric,
and the centroid does not correspond to the true sphere centre.
The apparent particle locations are too close together, which
explains contributions to $g(r)$ at distances less than $2r$.

\begin{figure}[h]
\begin{center}
\includegraphics[width=10cm,bb=0 0 800 410]{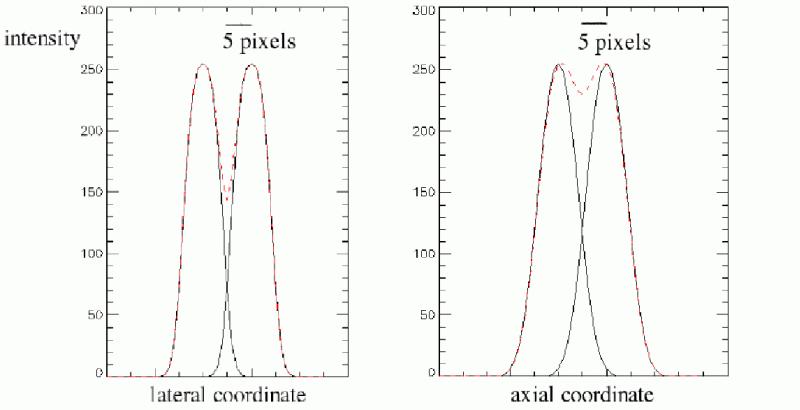}
\caption[The simulated intensity profile through two particles in
contact (lateral and axial).]{Calculated intensity profiles through
two particles in contact, both in lateral (left) and axial (right)
direction.\label{touchingparticles}}
\end{center}
\end{figure}

The intensity profiles calculated for two particles with diameter
$2~\mu$m at contact are shown in figure~\ref{touchingparticles}.
Between the particles the intensity drops to $57$\% of the peak
intensity in lateral direction (left). Due to the lower resolution
in axial direction, the intensity falls only to around $90$\%, if
the particles are touching in this direction (right). Depending on
the noise level, this might result in no detectable intensity drop
between these two particles.

The centroiding technique will thus become unreliable in dense
suspensions with a large fraction of touching or close particles.
The distance below which particles cannot be resolved any longer
depends on the conditions, but is typically below $1.3$ particle
diameters.

The fact that the image of a particle can be larger than the
particle itself is not specific to confocal microscopy. Since we
began investigating this effect, another group has suggested that
the corresponding effect for normal bright field microscopy has been
responsible for a significant controversy on so-called like-charge
attraction (e.g. \cite{Grier04,Baumgartl06}). It has been
demonstrated that the overlap of particle SSFs (our terminology) can
result in an apparently decreased or increased separation of
particles \cite{Baumgartl05}. The increased separation arises
because the intensity profile of a single particle is non-monotonic
for brightfield microscopy \cite{Hecht,Baumgartl05}, rather than the
(overly simplistic) monotonically decreasing form used here.  (Here
the Gaussian approximation to the PSF is notably inadequate.)

The problem of SSF overlap can be avoided by using so-called
core-shell particles  (for example, \cite{vanBlaaderen95}). They
consist of a fluorescent core and a non-fluorescent, but otherwise
identical shell. The shell is large enough that the SSFs of touching
particles do not significantly overlap.  Core-shell particles are,
however, more difficult to obtain and necessarily have fewer bright
pixels per particle, so that it may be necessary to oversample the
images to provide sufficient data to allow the centroiding procedure
to work.  Care must also be taken regarding polydispersity: the
cores can be polydisperse when the shells are not, and {\em vice
versa}.

\mod{We feel it is appropriate here to reiterate that despite the
limitations of centroiding, it is certainly an effective and robust
technique under suitable circumstances.  Its relative simplicity
commends its use where possible.  In particular, it is reliable in
dilute samples, where there is a small probability of SSFs
overlapping one another. Furthermore, although we have not dealt in
depth with polydisperse samples, the fact that centroiding relies
only on the particles being spherically symmetric suggests that it
ought to be applicable to these systems.}

\section{SSF refinement \label{SSFrefinement}}
To address some of the difficulties inherent in the centroid method,
we have developed an iterative technique which takes into account
the known shape of the SSF, using the `Refinement using the Measured
SSF' strategy (Section \ref{findbrightsandrefine}); we refer to this
as SSF refinement.

As before, the raw image is first filtered
(Sec.~\ref{dealingwithnoise}) and, as with the centroiding
technique, the image searched for local maxima to determine
candidate particle locations. Then, instead of taking the centroid
of the image intensity distribution within a window surrounding the
candidate particle location, an experimentally determined SSF is
fitted to the original image in the vicinity of the candidate
particle location. This is done, following similar work
\cite{Arndt66,Bobroff86}, with a chi-square test
\cite{BevingtonBook}.

\subsection{Determination of the Experimental SSF}
The average image of a large number of individual particles would
ideally provide an experimental SSF. This requires, however,
separate {\em and} slow-moving particles, which is difficult to
achieve. Furthermore, these particles should be part of the sample
of interest to have identical imaging conditions. Only in this case
does the SSF properly account for aberrations and imperfections such
as those caused by index mismatch between the sample and the
immersion oil, i.e. it genuinely reflects the system PSF. This can
even depend on, e.g., the depth in the sample. In contrast, a
different reference sample may not accurately reflect the image, and
thus SSF, of the particle in the sample of interest.

Therefore in most cases only an approximation to the ideal
experimental SSF can be determined. A region of appropriate size is
considered around each of the (nearest integer) candidate particle
locations as determined by the centroiding technique.  The
number-weighted mean of these images is then taken as the SSF,
$s_{exp}(x'',y'',z'')$. The resultant SSF is insensitive to a
number- or intensity-weighted average.  This average results in an
SSF which is broader than the `true' SSF, but this seems not to
affect significantly the quality of the obtained particle locations.

\subsection{Determining Particle Locations}
Having determined the SSF experimentally, we have established how
the vicinity of each particle ought to look. We can now refine the
(nearest integer) candidate particle locations. Within a window
around these locations, we search for the location which best
matches the SSF. The size of this window is chosen to match the
accuracy of the centroiding procedure, typically less than $2$
pixels in each direction: in the worst cases, the first peak of
$g(r)$ begins to rise at about $70$\% of one diameter, $2r$, i.e.
contacting particles are apparently about $0.3 \times 2r$ too close
and their locations are thus wrong by about $0.15 \times 2r$, or
about $2$ pixels, since the diameter corresponds to typically
$11$--$13$ pixels. For particularly noisy images, this value should
be increased.

The experimentally determined SSF, $s_{exp}(x'',y'',z'')$, is
expressed in terms of its own (integer) coordinates $x''$, $y''$ and
$z''$ with $0{\le} x''{\le} extent(0)$, $0{\le} y''{\le} extent(1)$
and $0{\le} z''{\le} extent(2)$. The detected image $I(x',y',z')$ is
also known only at integer locations.  The image of particle $i$ is
extracted from this: $I_i(x'',y'',z'')$. We have chosen to change to
the double prime coordinates which are measured within particle
$i$'s own window, allowing straightforward comparison with
$s_{exp}$; coordinates in this system are straightforwardly related
to the particle's global coordinate $(x_i', y_i', z_i')$. For a
meaningful comparison between $s_{exp}$ and $I$, they must be
normalised consistently while exploiting the full dynamic range.
This can be achieved by normalising the SSF to occupy the entire
range of greyscales and then scaling the image of each feature,
$I_i(x',y',z')$, such that its peak height matches that of the SSF.
Noise renders this an approximation, but seems not to significantly
affect the procedure.

The chi-square hypersurface of particle $i$ as a function of the
overlap coordinates $(j,k,l)$, is then:
\begin{displaymath}
\chi_i^2(j,k,l)=\sum_{x'',y'',z''}\frac
{\left[s_{exp}(x'',y'',z'')-I_i(x''+j,y''+k,z''+l)\right]^2}
{\sigma^2_{x'',y'',z''}},\end{displaymath} where the sum runs over
all $x''$, $y''$, and $z''$ (i.e. every pixel value in the SSF is
compared with every value in the current particle `image') and $j$,
$k$ and $l$ take the values $0\le j \le 2\delta_1$, $0 \le k \le
2\delta_2$ and $0\le l \le 2\delta_3$ with the iteration grid size
$(2\delta_1+1,2\delta_2+1,2\delta_3+1)$.\footnote{The grid
coordinates run $0\le j \le (2\delta_1+1)$ etc. in the SSF/particle
(`double prime') coordinate system, or $-\delta \le j \le \delta$ in
the global (`single prime') coordinate system.} (This makes it
impossible to extract features which lie within half of the
iteration grid size of the edge, as for the centroiding technique.)
The uncertainty associated with each point, $\sigma_{x'',y'',z''}$,
is assumed to be constant, although photon counting statistics
suggests an intensity dependence. Again, we obtain good results
despite this approximation.

\begin{figure}[h]
\begin{center}
\mbox{
\includegraphics[width=5cm,angle=0,bb=0 0 400 226]{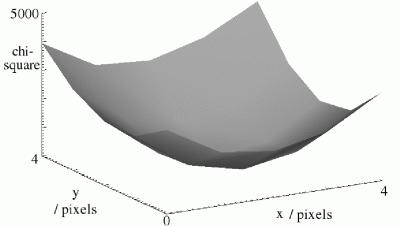} \qquad
\includegraphics[width=5cm,angle=0,bb=0 0 400 216]{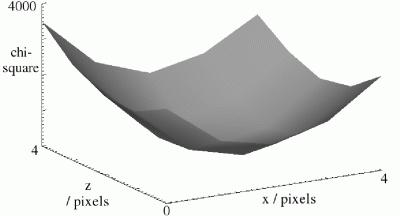}}
\caption[Two-dimensional projections of the Chi-square
hypersurface.]{Two-dimensional x-z slice through the chi-square
hypersurface for a randomly-chosen particle.
\label{chisquareonlattice2}}
\end{center}
\end{figure}

The chi-square values display a minimum near to the genuine particle
location (Fig.~\ref{chisquareonlattice2}), which, due to the
different resolution, is usually shallower in the axial ($z$) than
in the lateral ($x$, $y$) directions. Everything so far has involved
integer coordinates; for subpixel resolution, we need some form of
interpolation. To locate the minimum to sub-pixel accuracy, we
consider two interpolation schemes.

The first is to produce a fit to the SSF which is overlain on the
image and moved in any direction. This not only requires a fit
function, which is in general not available, but also leads to the
ambiguity that a variation in $\chi^2$ can either be due to a change
in the quality of the fit to the image, which is desired, or to a
change in the quality of the interpolation of the SSF.

The strategy we adopt here is different. We determine the minimum by
interpolating the $\chi^2$ surface in the three directions ($j$,
$k$, $l$) independently, typically using the five data points
closest to the minimum in each direction. By repeatedly
interpolating for a large number of points lying within half a pixel
of the minimum of the $\chi^2$ array, we obtain an array of
sub-pixel estimates to $\chi^2$. The lowest-valued entry in this
list gives the best correspondence between the measured SSF and the
image of the feature. A precision of one hundredth of a pixel is
more than necessary, but still computationally tolerable.  A crude
error estimate can also be made by estimating the error in the
interpolation; one popular interpolation algorithm, POLINT, does
this by default \cite[\S3.1]{NumRecipes}.

\subsection{Appraisal of the SSF refinement technique\label{SSFimprovement}}

\begin{figure}[h]
\begin{center}
\includegraphics[height=4.6cm,angle=0,bb=0 0 800 273]{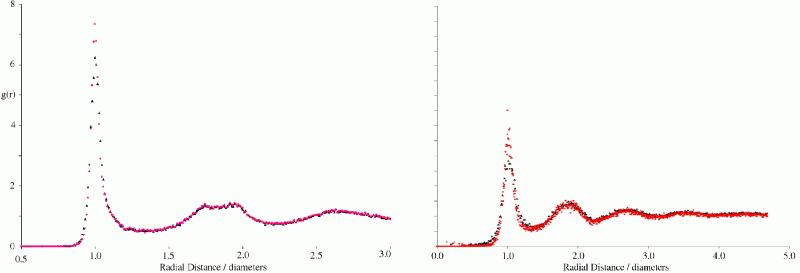}
\caption[The improvement in g(r) for a mediocre quality image of a
glassy sample.]{Comparison of the radial distribution functions
$g(r)$ as obtained by the SSF refinement technique (red circles) and
the centroiding method (black triangles) respectively. The analysis
is based on the same images as in figure~\ref{goodandbadglassyrdf};
a good quality (left) and mediocre quality (right) image of two
similar glassy samples (of the same system, $\Phi{\simeq}0.64$).
\label{badimageSSFrefineimprovement}}
\end{center}
\end{figure}

We illustrate the result of a SSF refinement using two examples
(Fig.~\ref{badimageSSFrefineimprovement}); further examples can be
found in the literature \cite{mythesis, mybridgespaper1}. They are
based on the previously used data set
(Fig.~\ref{goodandbadglassyrdf}). In the case of the high quality
image (Fig.~\ref{badimageSSFrefineimprovement} (left)), the obtained
$g(r)$ is reassuringly similar to the one from the centroiding
technique. The first peak begins to rise at very slightly higher
$r$, about $0.9$ diameters, which is reasonable, because the
polydispersity of the sample is about $5$\%. In addition to the
sharpening, the height of the first peak is increased by about
$20$\%, which indicates that even a relatively satisfactory analysis
can be improved by the SSF refinement technique.


The situation is different for an image of mediocre quality
(Fig.~\ref{badimageSSFrefineimprovement}, right). Here the $g(r)$
obtained by the centroiding technique is quite poor. In contrast,
the SSF refinement technique results in a $g(r)$ with a first peak
which is sharper, it begins to rise at about $0.8$ diameters (as
opposed to $0.6$ diameters for the $g(r)$ from the centroiding
technique), and higher, by more than $50$\%. We also notice the
occurrence of some noise at very small $r$; since $g(r)$ is
calculated by dividing by $r^3$, this can be caused by a very small
number of erroneous particle locations. (As we will explain below,
these can be identified.) The SSF refinement technique results, in
particular for sub-optimal data, in an improved quality of particle
location.

SSF refinement works partly because it searches neighbouring pixels,
that is, it does not assume that the centroid coordinates are
correct to the nearest pixel.  Noise in the data means that, despite
the noise filtering step, the brightest point of a particle's image
may not be the nearest to its true position.  This might also apply
to dilute suspensions with fast-moving particles \cite{Royall03}.
Since the centroiding procedure uses the brightest point as the
nearest-pixel estimate, this leads to an error of at least one pixel
in some cases.  An iterative version of the centroiding procedure,
should show some of the improvement we have found.  This requires
goodness of fit information for each individual particle, which we
provide here.

In $\chi^2$ we have a measure for the accuracy of the location of
each individual particle. We can thus not only compare the quality
of different images, but also individual particle locations. \mod{It
is important to remember that any particle location scheme relies on
a priori knowledge of the sizes and shapes of the particles.  The
less we know about any given particle (for example, how much its
radius differs from the mean), the less successful the particle
location is likely to be. This will be reflected in the $\chi^2$
value, which can be obtained on an individual particle basis.
Whether a poor $\chi^2$ value is due to a poorly-determined location
or a poor match between the particle image and the target SSF has
then to be investigated. This information} can be used when
calculating and interpreting structural properties, such as the
radial distribution function $g(r)$, based on the particle locations
(Sec. \ref{erroranalysis}).


\mod{Since SSF Refinement is more complex than centroiding, it
requires more processing time, by up to an order of magnitude.
However, the analysis can still reasonably be performed using
routines which were written in the IDL programming language without
particular attention to efficiency, and on a typical desktop
computer (AMD Athlon 3800+ 2.4Ghz, 2Gb RAM, running Windows XP). For
a typical image ($512\times512\times100$ voxels, containing about
$6400$ particles) and standard parameters
($extent{=}[15.0,15.0,13.0]$, iteration grid size
$\delta_1{=}\delta_2{=}\delta_3{=}2$ pixels) around $45$ seconds and
just over three minutes are required for the centroiding and SSF
Refinement, respectively.  (These times include the noise filtering,
which has to be done prior to both procedures, and in the example
given takes around half a minute.)  The error analysis, which will
be described in the following section, is of similar computational
complexity, but runs more slowly due to the slow routine used for
converting the $\chi^2$ value to a probability (Equation
\ref{errorequation}).}

\begin{figure}[h]
\begin{center}
\mbox{\includegraphics[width=8cm,angle=0,bb=0 0 600
321]{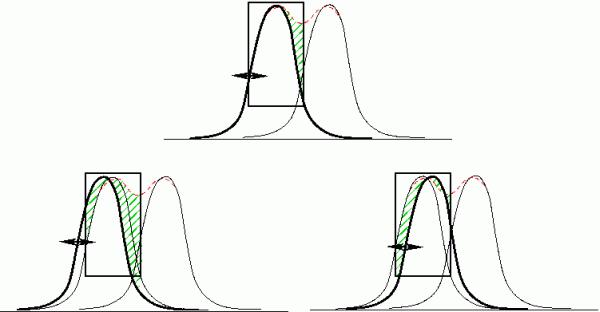}} \caption[The effect of overlapping SSFs on
chi-square.]{SSF refinement technique for two neighbouring
particles. See text for details. \label{SSFoverlapchisquare}}
\end{center}
\end{figure}

Despite its advantages, SSF refinement does not wholly overcome
the problem of close particles (Fig.~\ref{SSFoverlapchisquare}).
We consider three cases: First, even if the particle location is
determined correctly (top) and the SSF (thick line) is thus
aligned with the particle's image (thin lines), the overlap
with the neighbouring particle leads to a non-zero
contribution to $\chi^2$ (shaded area) in the window within
which it is calculated (rectange). This is not necessarily
important; the analysis relies only on $\chi^2$ having a
minimum at the particle location. Second, if the particle
location, and thus the SSF, is moved away from the
neighbouring particle (left), then the contribution to
$\chi^2$ (shaded area) is increased and this location
would thus quite rightly be rejected by the SSF
refinement procedure. Third, if the particle location
is moved closer to the neighbouring particle (right),
compared to the correct location the initially shaded
area now shrunk while a new shaded area
is created on the opposite, left side of the SSF. How
these two effects balance, and thus how $\chi^2$
changes, depends on the precise shape of the SSF
and the particle image. Nevertheless, this illustrates
that the minimum in $\chi^2$ might shift toward the
neighbouring particle. This is similar to the mentioned
effect of close particles on the centroiding technique,
although we expect it to be less significant for the SSF
refinement technique.

There are several possibilities how one could deal with this
problem. Knowledge of the SSF allows us to determine the intensity
profile for any set of particles and to fit several particle
locations simultaneously. However, this is computationally very
intensive. A less demanding approach would be to use $\chi^2$ on a
pixel level and develop a strategy to identify neighbouring
particles by their anisotropic $\chi^2$ (shaded areas in
Fig.~\ref{SSFoverlapchisquare}) and then either disregard these
pixels which are supposedly affected by a neighbouring particle or
use (the minimum in) their number to determine the particle
location.

\subsection{Error Analysis\label{erroranalysis}}
We showed that $\chi^2$ contains information on the reliability of
each particle coordinate. In addition, we can determine the
accuracy of the location in an arbitrary direction.  The shape of
the $\chi^2$ hypersurface reveals how sensitive this value is
to variations in the fit parameters, and these can be used to
infer the error in these quantities.

Recent papers have suggested a means of error determination based on
the mean variation between a particle's presumed stationary location
determined in a series of `identical' image captures
\cite{Dibble06,Savin05}, i.e. precision.  In contrast, we aim for a
measure which is both measured at the single particle level and
genuinely quantifies accuracy rather than precision.

First, we consider our expectation for the absolute value of
$\chi^2$. This depends not only on the uncertainty in each data
point, but also the number of degrees of freedom, $\nu$; it is usual
to define the reduced chi-square value $\chi_r^2=\chi^2/\nu$. This
value is expected to take the value $\chi_r^2=1$ if the data are
described well by the model, and the uncertainties
$\sigma_{x'',y'',z''}$ are representative \cite{BevingtonBook}.

To obtain an estimate for the errors in the fit parameters, i.e. the
particle locations, we consider the shape of the $\chi^2$
hypersurface. There are established means of estimating the error in
the position of the minimum
\cite{BevingtonBook,Arndt66,Bobroff86,NumRecipes,Lampton76,Pedersen97}.
If we neglect any error in sampling the surface, that is, assume
perfect interpolation, then the error is found from the curvature of
the surface. It is customary to take the error in a parameter as
being the change in its value for which $\chi_r^2$ has increased by
unity, $\Delta\chi_r^2{=}1$. This assumes that the experimentally
determined minimum of $\chi_r^2$ is unity, $\min\{\chi_r^2\}{=}1$.
If the model is appropriate but the noise is unknown, then the
condition can be recast: the error in $\chi_r^2$ is found when it
has risen by a value equal to $\min\{ \chi_r^2\}$, i.e.
$\Delta\chi_r^2{=}\min\{\chi_r^2\}$. Crucially, it is therefore not
necessary to estimate the uncertainty in each pixel intensity value.
Moreover, since $\chi^2$ and $\chi_r$ are related by the degrees of
freedom, $\nu$, which is constant within a given experiment, we can
continue using $\chi^2$.

An increase $\Delta\chi^2{=}\min\{\chi^2\}$ gives the probability of
finding the position between the genuine centre and that location of
$68.3$\%, that is, at the one $\sigma$ confidence level.  This
follows from the chi-square cumulative distribution \cite{Cramer46}:

\begin{equation}
F(\Delta\chi^2; \nu) =
\frac{\int_0^{\Delta\chi^2/2}t^{\left(\nu/2\right)-1}e^{-t}dt}{\int_0^\infty
t^{\left(\nu/2\right)-1}e^{-t}dt}=\frac{\gamma\left(\nu/2,\Delta\chi^2/2\right)}{\Gamma\left(
\nu/2 \right)}.\label{errorequation}\end{equation} Values of
$F(\Delta\chi^2; \nu)$ are tabulated in
reference~\cite[\S14.5]{NumRecipes}.

Using this relation, we can calculate the probability that the
particle centre lies between the minimum of the
$\chi^2$-hypersurface and any point on that surface, i.e. for any
$\Delta\chi^2$. The $\chi^2$-hypersurface can therefore be mapped to
a probability surface.

\begin{figure}[htp]
\begin{center}
\mbox{
\includegraphics[height=5cm,bb=0 0 400 263]{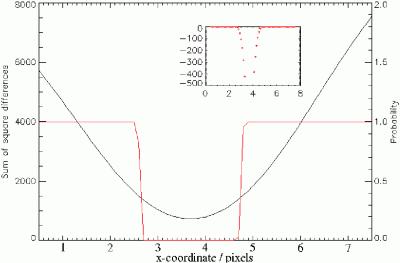}}
\mbox{
\includegraphics[height=5cm,bb=0 0 400 263]{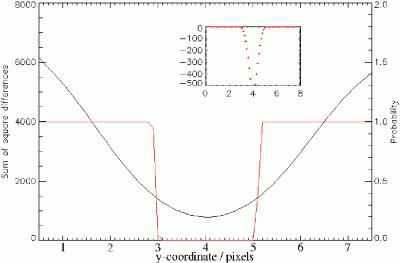}}
\mbox{
\includegraphics[height=5cm,bb=0 0 400 263]{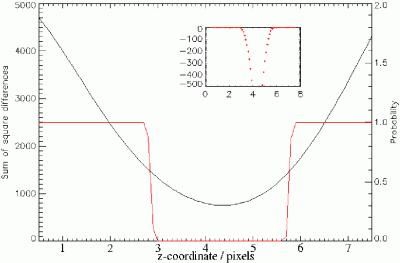}}
\caption{Variation of $n\chi^2/\min{\chi^2}$ (black curve, left-hand
axis) in the vicinity of its minimum, i.e. the most likely particle
location, in $x$-, $y$- and $z$-directions (from top). The centre of
these images correspond to the location determined by the centroid
analysis. Also shown is the cumulative probability (red, right-hand
axis) with the inset showing the logarithm of this probability
(Importantly, the probability is calculated using the minimum of the
whole $\chi^2$ surface, which in general differs from the minimum
through any particular cut). The pixel pitch was laterally
$0.16~\mu$m and axially $0.20~\mu$m. Furthermore, $\chi^2$ is only
known at integer pixels, while sub-pixel values are obtained by a
one dimensional polynomial interpolation. \label{probdists}}
\end{center}
\end{figure}

A probability distribution created from the $\chi^2$-hypersurface is
shown in figure~\ref{probdists} (black curves, left-hand axis) as
cuts through a particle in $x$-, $y$- and $z$-direction (from top).
Also shown (red) is the cumulative probability that a particle
centre will be found somewhere between the minimum of $\chi^2$ and
that location.  We can hence infer an error bound; the width of the
region of the cumulative probability curve which has height less
than $0.683$, the `$1\sigma$ point'.  It is about $180$~nm in the
lateral direction, and about $300$~nm in axial direction.

\begin{figure}[h]
\begin{center}

\includegraphics[width=8cm,bb=0 0 600 815]{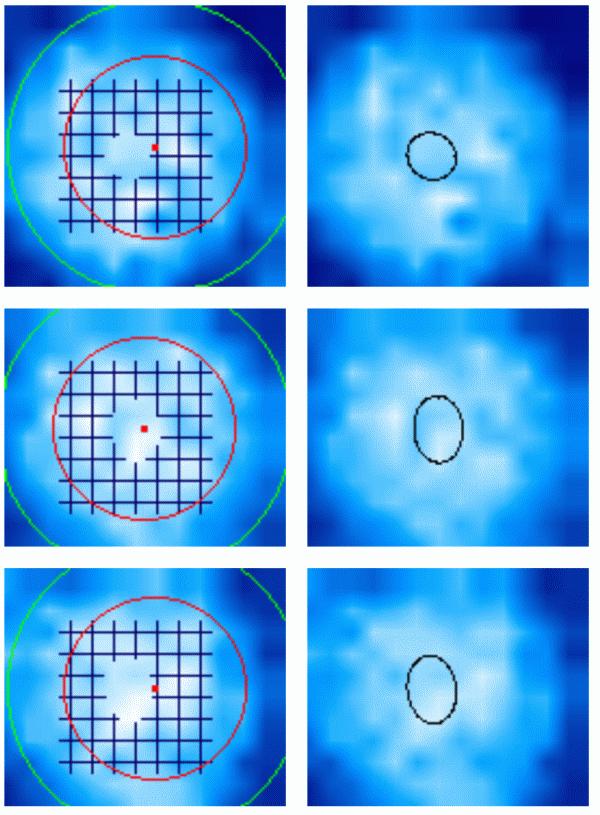}
\caption{$xy$-, $yz$- and $xz$-slices (from top) through the
probability distribution, overlain onto the original confocal data
(shown as a blue and white background). The centre as determined by
the centroid analysis is shown as a red point (left), and, centred
on the point, circles indicating the size of the particle as
determined by the centroid analysis (red) and dynamic light
scattering (green) respectively. The black grid represents those
points where the probability lies above the $1 \sigma$ value. The
locus of points at the $1 \sigma$ level are represented by ellipses
on the right. (See text for details.)
 \label{overlainpointsandellipses}}
\end{center}
\end{figure}
\begin{figure}[htp]
\begin{center}
\includegraphics[width=10cm,bb=0 0 600 270]{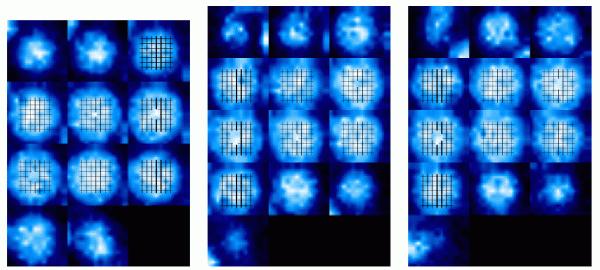}

\caption{Series of $xy$-, $xz$- and $yz$-slices (from left) through
the probability distribution of the same particle as discussed in
Figure \ref{overlainpointsandellipses}. \label{overlaingrids}}
\end{center}
\end{figure}
\begin{figure}[htp]
\begin{center}
\includegraphics[width=8cm,bb=0 0 600 568]{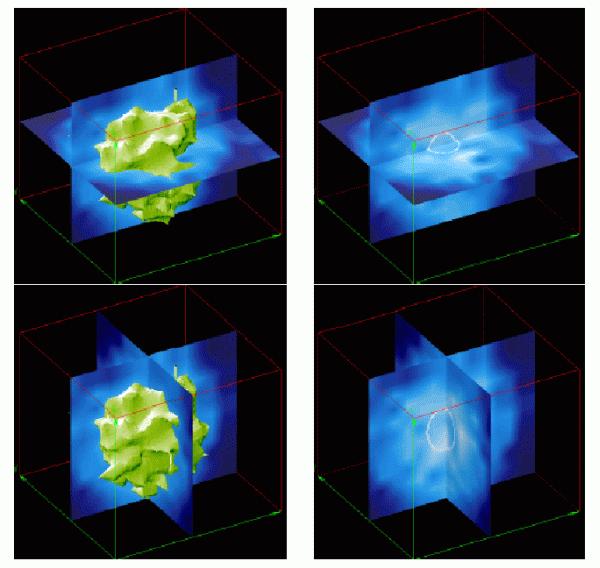}

\caption{Three-dimensional representation of a particle (left) and
its error bound (right, white ellipses) which represents the
$1\sigma$ confidence limit on the particle location and thus
delimits the region where the particle is most likely found. The
particle is represented by an isosurface (contour) at greylevel 200.
The top images show $y$- and $z$-slices, while the bottom images
show $x$- and $y$-slices. \label{slicer}}
\end{center}
\end{figure}

We can represent these more informatively.  Figure
\ref{overlainpointsandellipses} shows the entire particle as viewed
in the $xy$-, $xz$- and $yz$-planes (top to bottom).  The results
from the centroid procedure are overlain: the centre and an
indication of the particle size are shown (left, red dot and
circle), as well as the particle size as determined by light
scattering (green circle; both circles centred on centroid
location). The black grid represents those points where the
cumulative probability $p$ is greater than the $\sigma$ value ($p
\ge 0.683$) (the probabilities, according to the prescription given
above, are calculated on a grid), as this highlights (in the middle)
the region below the $1 \sigma$ value, which defines the likely
particle location. For clarity, we have extracted the locations
which denote the perimeter of the gap in the grid, and fitted
ellipses through these (right). These represent the $1\sigma$
confidence level for the particle locations. These are not
symmetric, in particular, they are elongated along the axial
direction, as we may have expected.
Furthermore, although these look believable when compared with the
original confocal image (the blue-white background), neighbouring
slices also contribute to the determined location, making subjective
conclusions difficult. A series of slices is shown in
figure~\ref{overlaingrids}; these are cuts through $xy$-, $xz$-, and
$yz$-directions from top to bottom. They are combined in a
three-dimensional representation in figure~\ref{slicer}, which
summarizes the results obtained by the SSF refinement technique.
Ellipses representing the $1 \sigma$ confidence limit on the
particle location (right) is compared to the contour surface based
on slices through the raw image of a particle (left).



\begin{figure}[htp]
\begin{center}
\mbox{
\includegraphics[height=5cm,bb=0 0 400 250]{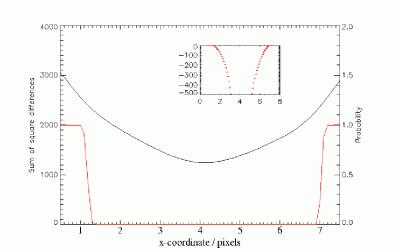}}
\mbox{
\includegraphics[height=5cm,bb=0 0 400 250]{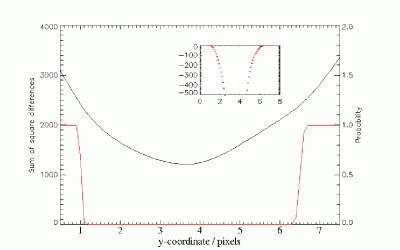}}
\mbox{
\includegraphics[height=5cm,bb=0 0 400 250]{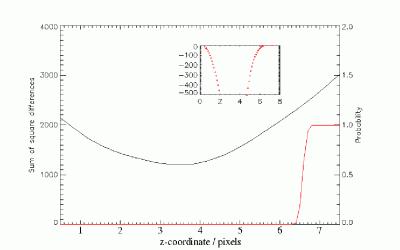}}

\caption{Same information as in Figure \ref{probdists}, but for
lower image quality: Variation of $\chi^2$ (black curve, left-hand
axis) and cumulative probability (red) in the vicinity of the
minimum, i.e. the most likely particle location, in $x$-, $y$- and
$z$-directions (from top). The pixel pitch was laterally $0.13~\mu$m
and axially $0.20~\mu$m. \label{probdistsbad}}
\end{center}
\end{figure}

\begin{figure}[h]
\begin{center}
\includegraphics[width=8cm,bb=0 0 800 265]{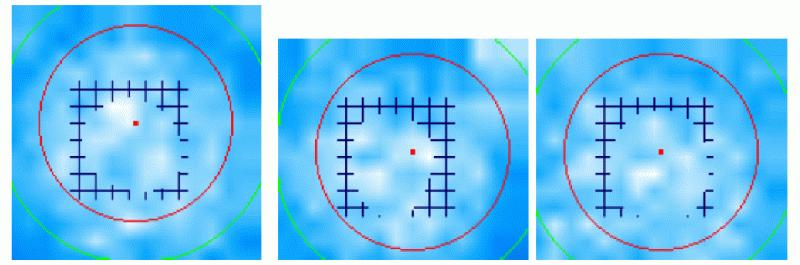}
\caption{Same information as on the left in
figure~\ref{overlainpointsandellipses}, but for a lower quality
image. \label{overlainpointsandellipsesbad}}
\end{center}
\end{figure}

\begin{figure}[htp]
\begin{center}
\includegraphics[width=10cm,bb=0 0 600 276]{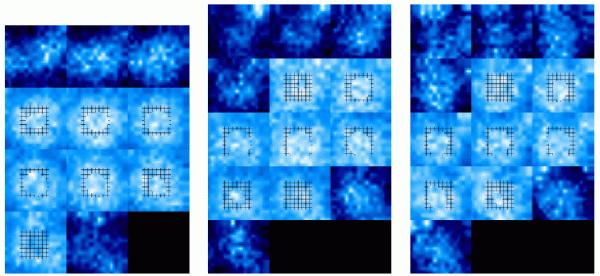}
\caption{Same information as in figure~\ref{overlaingrids}, but for
a lower quality image. \label{overlaingridsbad}}
\end{center}
\end{figure}

For comparison we now present the corresponding results for an image
of lower quality, but similar sample, PMMA spheres with radius
$2.18~\mu$m in {\em cis}-decalin (Figs.~\ref{probdistsbad} --
\ref{overlaingridsbad}). The error bounds calculated based on the
$\chi^2$ surface (Fig.~\ref{probdistsbad}) are about $360$--$380$~nm
in lateral and about $620$~nm in axial direction. These are about
double the values found for the good quality image
(Fig.~\ref{probdists}). The axial uncertainty is determined from one
direction only (bottom), because the probability does not rise
sufficiently within the window studied (though the window size can
be increased, care must be taken not to include contributions from
neighbouring particles). The lower image quality is also reflected
in the larger regions of likely particle locations, quantified in
the $1 \sigma$ confidence level for the particle locations
(Figs.~\ref{overlainpointsandellipsesbad}
and~\ref{overlaingridsbad}), which are significantly larger than for
the good quality image (Figs.~\ref{overlainpointsandellipses}
and~\ref{overlaingrids}).


The different curvatures of the $\chi^2$-hypersurface for the
good (Fig.~\ref{probdists}) and low (Fig.~\ref{probdistsbad})
quality image contain the important information which
allows us to compare the accuracy of the coordinates extracted
from similar systems. When comparing systems which are
optically different (e.g. \cite{Conrad05}), a measure similar to
this one can thus be used.

On a quantitative level, the error bounds are rather large compared
with those cited by other studies; about $30$--$50$~nm. The
systematic errors probably explain this:  The error we have
quantified above is the inherent error arising from the noise, SSF
overlap, and other sources of error. However, to assume that we can
use $\Delta\chi^2=\min\{\chi^2\}$ as the error criterion rather than
the usual unity is only justified if the data are well-modelled by
the SSF.  When this is not true, the relative increase in $\chi^2$
is overestimated, and correspondingly the error overestimated. In
this paper the experimental SSF is determined as an average which is
necessarily broader than its true form; this certainly gives rise to
a systematic error. Nonetheless, we feel that the error determined
as described above is a genuine and informative measure of the error
in the particle locations. It is indicative of the shape of the
confidence contours, and gives a compelling comparison of the
accuracy between different particles.

\begin{figure}[htp]
\begin{center}
\includegraphics[height=5cm,bb=0 0 600 375]{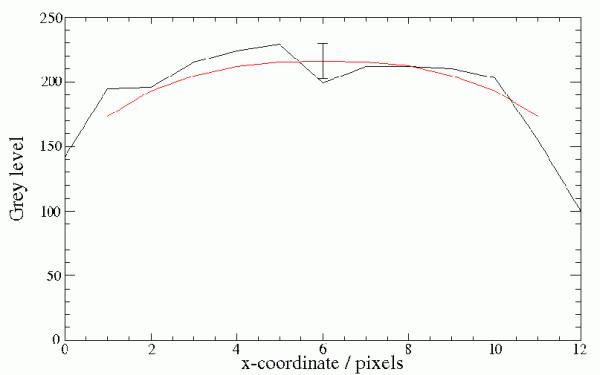}
\caption{One-dimensional cut through the original raw data
corresponding to the particle studied in
figures~\ref{probdists}--\ref{slicer} (black).  Also shown is the
same cut through the calculated SSF (red), and a suggestive error
bar exactly one half the size implied by the analysis described in
the text.  Note that these latter two are aligned with the centre of
the graph, whereas the real data are sampled at integer locations
and are therefore necessarily not (though all are nearly aligned).
\label{impliednoise}}
\end{center}
\end{figure}

The foregoing analysis was performed without the need for an
estimate of $\sigma_i$, the error on each pixel's intensity.  We
can, however, infer what the mean of this value must be, supposing
the model is correct, i.e. the extracted SSF is representative. For
the particle shown in Figures \ref{probdists}-\ref{slicer}, the
minimum of the $\chi^2$ surface was about $750$ per pixel.
Neglecting any intensity dependence of $\sigma$, for example due to
photon-counting related noise, this gives an estimate of the mean
noise in each pixel $\overline{\sigma_i}{=}\sqrt{750}\simeq 27$.
Figure \ref{impliednoise} shows a one-dimensional cut through the
raw data for the particle shown in Figures
\ref{probdists}-\ref{slicer}, along with an error bar which to the
eye is approximately the correct size.  Its size is $27$ in total,
about half the size implied by the above calculation, revealing as
expected that systematic errors contribute to the $\chi^2$ surface.

\begin{figure}[h]
\begin{center}
\includegraphics[width=10cm,bb=0 0 400 277]{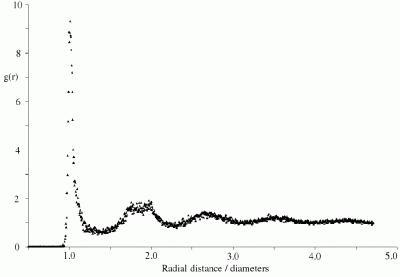}
\caption[The improvement in g(r) caused by discarding any particles
with a poor $\chi^2$ value.]{Radial distribution function $g(r)$
obtained by discarding particles with a poor $\chi^2$ value.
\label{bestchi}}
\end{center}
\end{figure}

The information on the reliability of particle locations is
furthermore very useful in those cases where it is not necessary to
consider all of the particles in a sample. Then only those particles
can be taken into account which are known, via $\chi^2$, to be
reliably located. However, care must be taken when discarding
information; a discarded particle may be accurately located but have
a poor $\chi^2$ value for another reason, for example, if it has a
different size. There is, for example, a clear improvement in $g(r)$
caused by considering only particles with better than a certain
$\chi^2$ value (Fig.~\ref{bestchi}, showing the same sample as on
the left in Fig.~\ref{goodandbadglassyrdf}). Since these particles
match the SSF best, they must both be located to high precision and
similar to the expected shape.


In some cases, it may be of interest to specify an SSF which
corresponds to a sub-population within the sample, for instance
relatively small particles. SSF refinement with $\chi^2$
discrimination would then preferentially select a sub-population, in
this case by ignoring larger particles.  Due to the inherent
variability in the images of particles, this is realistic only for
quite differently-sized particle.

Moreover, we can apply different SSFs representing different shapes
or sizes of particle and, in particular, one may be interested in
replacing the SSF with the PSF.  We noted earlier the similarity
between deconvolution of instances of the SSF and those of the PSF.
Thinking along these lines suggests using the procedure for finding
instances of the SSF for locating instances of the PSF.  If the
system under study consists of point-like visible components, then
the SSF of each is identical to the PSF.  In a system matching this
description, for example one comprising particles with a very small
fluorescent core, the SSFs would not overlap and SSF refinement
would be accurate without approximation to within the uncertainty of
the imaging system.

\section{Conclusion}
Confocal microscopy has become a powerful tool for studying
colloidal systems, in particular to obtain quantitative information
on the level of individual particles. Here we have reviewed and
extended methods for particle location in colloidal systems with a
special emphasis on dense systems. We have considered the different
steps finally leading to reliable particle coordinates and the
associated errors; from optimum image formation to the initial
finding of candidate particle locations and the subsequent
refinement by both centroiding and SSF refinement, and ultimately to
a detailed, novel error analysis which provides confidence regions
for individual particles.

Particular emphasis was put on image analysis, where we focused on
two important methods. First, we reviewed the centroiding technique,
which is widely used in colloid science \cite{Crocker96}, but have
maintained generality by outlining some other feature location
strategies. Second, we have introduced a technique previously
unreported in colloidal studies: SSF refinement. This not only
optimises particle coordinates, but also allows for a detailed error
analysis by means of the $\chi^2$ measure.  It permits a
quantitative comparison between different systems and, furthermore,
is sufficiently general to be useful in most feature extraction
applications. Moreover, we have demonstrated how to locate particles
reliably, albeit not necessary precisely, under adverse imaging
conditions, such as dense colloidal suspensions where features may
overlap. The confidence intervals based on a  $\chi^2$ analysis are
instrumental in achieving this.

\addcontentsline{toc}{section}{Acknowledgments}
\section*{Acknowledgments}
We are grateful to the UK Engineering and Physical Sciences Research
Council (EPSRC), the Deutsche Forschungsgemeinshaft (DFG)
(Collabarative Research Centre SFB-TR6, Project Section C7) and to
Rhodia Research and Technology (Rhodia -- Centre de Recherches et
Technologies d'Aubervilliers) for support, to Andrew Schofield in
The University of Edinburgh School of Physics for supplying the
particles and for helpful discussions, and to Eric Weeks for
providing the three-dimensional version of the centroid-based
particle tracking software \cite{WeeksTrackingWeb}. We thank Jan
Skov Pedersen, Wilson Poon, and Mark Haw for helpful discussions.

\bibliographystyle{elsart-num}
\bibliography{JenkinsEgelhaafACIS2007_tocondmat}

\newpage
\clearpage

\newpage

\newpage

\newpage

\newpage
\newpage

\newpage\clearpage

\newpage\clearpage

\newpage
\newpage

\newpage

\newpage
\newpage

\newpage

\newpage

\newpage

\newpage

\newpage
\newpage

\newpage

\newpage

\newpage
\newpage
\newpage

\newpage

\newpage

\newpage

\newpage
\newpage\clearpage

\end{document}